\documentclass[aps,twocolumn,10pt,prc,floatfix,showpacs,preprintnumbers,amsmath,amssymb,nofootinbib,superscriptaddress]{revtex4-1}

\usepackage[normalem]{ulem}
\usepackage{wasysym}
\usepackage{color}
\usepackage{graphicx}
\usepackage{subfig}
\usepackage{dcolumn}    % Align table columns on decimal point
\usepackage{multirow, booktabs }
\usepackage{comment}
\usepackage{soul}

\usepackage[format=plain,labelsep=period]{caption}
\usepackage{physics, mathtools}   %, mathabx}
\usepackage[english]{babel}

\usepackage{amsmath,amsfonts,amssymb,amsthm}  
\usepackage{latexsym}
\usepackage{bbold}
\usepackage{feynmp-auto}
\usepackage{subcaption}
\usepackage{graphicx}
\usepackage[utf8]{inputenc}

\usepackage{tikz}
\usetikzlibrary{decorations.pathreplacing}

\usepackage{bigstrut}
\usepackage{CJK}
\usepackage[pdfstartview=FitH,
            CJKbookmarks=true,
            bookmarksnumbered=true,
            bookmarksopen=true,
            colorlinks,
            pdfborder=001,
            linkcolor=blue,
            anchorcolor=blue,
            citecolor=blue,
            urlcolor=blue,
            ]{hyperref}

\usepackage{cleveref}

\begin{document}

\newcommand{\IdentityMat}{\mathbb{1}}
\newcommand{\varQ}{\mathbf{q}}
\newcommand{\varK}{\mathbf{k}}
\newcommand{\varX}{\mathbf{x}}
\newcommand{\varR}{\mathbf{r}}
\newcommand{\Sch}{ Schr\"{o}dinger }
\newcommand{\coeffSch}{-\frac{\hbar^2}{2m}}
\newcommand{\etal}{\textit{et al.} }
\newcommand{\secondfdv}[3]{ \frac{\delta^2 {#1}}{\delta #2 \, \delta #3} }
\newcommand{\RomanNumeralCaps}[1]
    {\MakeUppercase{\romannumeral #1}}

\newcommand{\FM}[1]{{\color{magenta} #1}}
\newcommand{\can}[1]{{\color{red}\sout{#1}}}
\newcommand{\SN}[1]{{\color{orange} #1}}
\newcommand{\SNo}[1]{{\color{orange}\sout{#1}}}
\newcommand{\WG}[1]{{\color{cyan} #1}}

\author{F. Marino}
\email{frmarino@uni-mainz.de}
\affiliation{Institut f\"{u}r Kernphysik and PRISMA+ Cluster of Excellence, Johannes Gutenberg-Universit\"{a}t Mainz, 55128 Mainz, Germany}
%\affiliation{Dipartimento di Fisica ``Aldo Pontremoli'', Universit\`a degli Studi di Milano, 20133 Milano, Italy}
%\affiliation{INFN,  Sezione di Milano, 20133 Milano, Italy}

\author{W. G. Jiang}
\affiliation{Institut f\"{u}r Kernphysik and PRISMA+ Cluster of Excellence, Johannes Gutenberg-Universit\"{a}t Mainz, 55128 Mainz, Germany}
\affiliation{Mainz Institute for Theoretical Physics, Johannes Gutenberg-Universit\"{a}t, 55128 Mainz, Germany}

\author{S. J. Novario}
\affiliation{Department of Physics, Washington University in Saint Louis, Saint Louis, MO
63130, USA}

%\author{C. Barbieri}
%\affiliation{Dipartimento di Fisica ``Aldo Pontremoli'', Universit\`a degli Studi di Milano, 20133 Milano, Italy}
%\affiliation{INFN,  Sezione di Milano, 20133 Milano, Italy}

%\author{G.~Col\`{o}}
%\affiliation{Dipartimento di Fisica ``Aldo Pontremoli'', Universit\`a degli Studi di Milano, 20133 Milano, Italy}
%\affiliation{INFN,  Sezione di Milano, 20133 Milano, Italy}

%\author{G. Hagen}
%\affiliation{Physics Division, Oak Ridge National Laboratory, Oak Ridge, TN 37831, USA}
%\affiliation{Department of Physics and Astronomy, University of Tennessee, Knoxville, TN 37996, USA}

%\author{T. Papenbrock}
%\affiliation{Department of Physics and Astronomy, University of Tennessee, Knoxville, TN 37996, USA}
%\affiliation{Physics Division, Oak Ridge National Laboratory, Oak Ridge, TN 37831, USA}

\title{
Diagrammatic \textit{ab initio} methods for infinite nuclear matter with modern chiral interactions
}

\begin{abstract}
A comparative study of the equation of state for pure neutron matter and symmetric nuclear matter is presented using three \textit{ab initio} methods based on diagrammatic expansions: coupled-cluster theory, self-consistent Green's functions, and many-body perturbation theory. We critically evaluate these methods by employing different chiral potentials at next-to-next-to-leading-order--all of which include both two- and three-nucleon contributions--and by exploring various many-body truncations.
Our investigation yields highly precise results for pure neutron matter and robust predictions for symmetric nuclear matter, particularly with soft interactions.
Moreover, the new calculations demonstrate that the $\rm{ NNLO_{sat} }(450)$ and $\Delta \rm{ NNLO_{go} }(394)$ potentials are consistent with the empirical constraints on the saturation point of symmetric nuclear matter.
Additionally, this benchmark study reveals that diagrammatic expansions with similar architectures lead to consistent many-body correlations, even when applied across different methods. This consistency underscores the robustness of the diagrammatic approach in capturing the essential physics of nucleonic systems.

\end{abstract}

\maketitle

\section{Introduction}

Infinite nuclear matter serves as an idealized and fundamental concept in nuclear physics, describing an extended, homogeneous system of strongly interacting nucleons. 
This theoretical construct is vital for understanding the properties of nuclear systems under extreme conditions, such as those found in the cores of neutron stars~\cite{Burgio2021,HaenselNeutronStars,Sumiyoshi2023,chatziioannou2024neutronstarsdensematter}, during supernova explosions, and in other dynamic processes involving the formation and evolution of compact celestial bodies. 
The essential property of nuclear matter is its equation of state (EOS), which, at zero temperature, describes the energy per particle as a function of the nucleonic density.

Comparing the theoretical predictions for the EOS with the empirical constraints from experimental nuclear physics and astrophysical observations~\cite{Watts2016,Baiotti2019,Gorda2022} provides insights into our understanding of nuclear interactions~\cite{Drischler2021Review,Huth2022}.
For example, potentials rooted in chiral effective field theory ($\chi$EFT) have long struggled to accurately reproduce the bulk properties of finite nuclei and the saturation point of symmetric nuclear matter (SNM) with the same underlying nuclear potential~\cite{Simonis2017,MACHLEIDT2024104117,Sammarruca2020}. However, consistency with the empirical constraints is now possible with more recent models~\cite{NNLOsat,DeltaGo2020,Drischler2019,Jiang2024,PbAbInitio}.

While pure neutron matter (PNM) is a relatively weakly correlated system at nuclear densities up to saturation density $\rho_{0}\approx 0.16\,\rm{fm}^{-3}$ and above~\cite{Gandolfi2015} (the uncertainties of which are predominantly related to the interaction, see e.g.~\cite{Piarulli2020,Lovato2022PNM,Drischler2020}), the nuclear force is rather strong in the $T=0$ isospin channel. 
Advanced many-body methods~\cite{Hergert2020,computational_nuclear} are thus essential to draw reliable estimates of the EOS and exploit the coinciding developments in \textit{ab initio} chiral potentials~\cite{Piarulli2020Delta,Machleidt2016,MACHLEIDT2024104117,Epelbaum2024}.

A variety of \textit{ab initio} techniques have been applied to determine the nuclear matter EOS. Among others, we mention Br\"{u}ckner-Hartree-Fock ~\cite{Burgio2021}, finite-temperature Green's functions~\cite{Rios2020,Carbone2020,Carbone2014,Carbone2013Sym}, and frameworks based on many-body perturbation theory (MBPT) ~\cite{Drischler2019,Drischler2021Review,Keller2023,Tichai2020Mbpt}.
Methods of the Quantum Monte Carlo family, such as Auxiliary field diffusion Monte Carlo (AFDMC)~\cite{Lynn2019,Gandolfi2015,Piarulli2020} and configuration-interaction Monte Carlo (CIMC)~\cite{Roggero2014CIMC,Arthuis2023}, have been used extensively for PNM, but feature few applications to SNM~\cite{Lonardoni2020LocalChiral,Gandolfi2014snm,Marino2021}.

Over the last decade, coupled-cluster (CC) theory, a popular and accurate method in both quantum chemistry~\cite{ShavittBartlett,Bartlett2007} and nuclear physics~\cite{Hagen2014Review}, has been applied to nuclear matter~\cite{Hagen2014,Baardsen2013}, mostly in conjunction with the interactions developed by the G\"{o}teborg and Oak Ridge groups~\cite{NNLOopt,NNLOsat,Ekstrom2018Delta,DeltaGo2020,PbAbInitio}, such as $\rm{ NNLO_{sat} }$~\cite{NNLOsat} and the delta-full models $\Delta \rm{ NNLO }$~\cite{Ekstrom2018Delta} and $\Delta \rm{ NNLO_{go} }$~\cite{DeltaGo2020}.
The CC method truncated at the level of doublet amplitudes with the addition of perturbative triples corrections [CCD(T)] recovers a large fraction of the correlation energy at a polynomially-scaling computational cost and is considered one of the best-performing \textit{ab initio} techniques for ground state (g.s.) energies.

At the same time, a self-consistent Green's functions (SCGF) method~\cite{Soma2020,Barbieri2004} based on the algebraic diagrammatic construction (ADC) approximation for the self-energy~\cite{Soma2020,Schirmer2018,Raimondi2017,Barbieri2022Gorkov}, which has been a state-of-the-art approach for finite nuclei for years (see e.g.~\cite{Soma2020Chiral,Arthuis_2020,Soma2014Chains,Raimondi2019}), has been proposed for infinite nuclear matter~\cite{Barbieri2017,McilroyChristopher2020Sgfs,Marino2024,MarinoPhdThesis,Marino2024Qnp}.
ADC-SCGF is a versatile tool for zero-temperature nuclear matter, which in a single calculation gives access to the total energy as well as single-particle (s.p.) properties such as the momentum distribution. 

CC and ADC-SCGF bear many similarities, as they are both non-perturbative techniques, where the g.s. is determined as the solution to a set of equations that involve the resummation of classes of diagrams to all orders in perturbation theory~\cite{Hagen2014Review,Soma2020}.
Also, they adopt the same scheme for simulating the infinite system, which is modeled by a finite number of nucleons in a box, subject to periodic boundary conditions (PBCs) and a finite model space~\cite{Hagen2014,Marino2024}.
Moreover, since they are formulated in momentum space, they can easily handle non-local chiral potentials, thus avoiding a known difficulty of coordinate-space techniques such as AFDMC~\cite{Piarulli2020Delta}.
It is now timely then, to perform a comparison between these two approaches, following a tradition of benchmark studies in infinite matter~\cite{Baldo2012,computational_nuclear,Piarulli2020,Lovato2022PNM} and finite nuclei~\cite{Hergert2016,Hergert2020,Scalesi2024}.

In addition to CC and SCGF, a third diagrammatic method is considered: MBPT at third order.
It is known~\cite{Drischler2021Review} that with soft chiral interactions, MBPT(3) provides a reasonable approximation to the nuclear matter energies at a smaller cost than the more sophisticated non-perturbative techniques.
Striving to make the comparison as consistent as possible, a new implementation of MBPT(3) in a finite model space has been developed, which uses the same discretization procedure as CC and ADC-SCGF, and differs from other studies, see e.g.~\cite{Hebeler2010,Drischler2019}, where energies are evaluated in the continuum limit.

The purpose of this work is to provide a detailed benchmark for nuclear matter calculations of both PNM and SNM at zero temperature between three independent \textit{ab initio} methods: CC, ADC-SCGF, and MBPT(3).
These approaches are reviewed in Sec.~\ref{sec: methods}.
Then, detailed numerical results are presented in Sec.~\ref{sec: results}.
Realistic nuclear forces at next-to-next-to-leading order (NNLO) in the chiral expansion, which comprise both nucleon-nucleon (NN) and three-nucleon (3N) interactions are employed.
We investigate different many-body truncation schemes and potentials to gauge the uncertainties that are carried by the approximations inherent to each of these methods and their sensitivity to the regularization cutoff of the interactions.
In addition, new sets of predictions for the nuclear matter saturation point of the G\"{o}teborg-Oak Ridge interactions are presented.
Finally, conclusions and perspectives of this paper are reported in Sec.~\ref{sec: conclusions}.

\section{Methods}
\label{sec: methods}

All the techniques presented in this section are based on a finite model space where the homogeneous system is simulated using a finite number of nucleons $A$ in a box of size $L$ and volume $L^3$, such that they are related to the number density $\rho$ by $\rho=A/L^3$, see e.g.~\cite{Marino2023,LietzCompNucl,Hagen2014}.
Periodic boundary conditions (PBCs) are imposed; hence, momentum states are quantized in units of $2\pi/L$, namely
\begin{align}
    \label{eq: sp mom state}
    \mathbf{k} = \frac{2\pi}{L} \mathbf{n},\qquad n_i=0,\pm 1, \pm 2 ...
\end{align}
for $i=x,y,z$.
It is convenient to choose $A$ to correspond to a closed-shell configuration in momentum space. Throughout this paper, $N=66$ neutrons and $A=132$ nucleons are employed in PNM and SNM, respectively. These ``magic numbers'' allow to minimize the finite-size effects~\cite{Hagen2014,LietzCompNucl}. 
In SCGF and MBPT, the s.p. basis is chosen so as to include all states \eqref{eq: sp mom state} within a sphere of squared radius $N^2_{max} = 25$ such that $n_x^2+n_y^2+n_z^2 \le 25$.
For CC, a cubic truncation is adopted, with $\abs{n_i} \le 4$.
We have verified that the results are converged with respect to the model space dimension for both types of truncations.

We utilize both delta-less and delta-full forces in this work, which are all truncated at NNLO in the chiral expansion and are composed of both NN and 3N interactions (see e.g.~\cite{Epelbaum2006Review,Machleidt2011,Piarulli2020Delta,MACHLEIDT2024104117}).
In particular, we employ $\rm{ NNLO_{sat} }$~\cite{NNLOsat}, the $\Delta \rm{NNLO}$ models of Ref.~\cite{Ekstrom2018Delta} and the $\Delta \rm{NNLO_{go}}$ potentials~\cite{DeltaGo2020}.
All these interactions make use of non-local regulators in both the NN and 3N sectors, whose cutoff (in $\rm{ MeV/c }$) is noted in parentheses, e.g. $\rm{ NNLO_{sat} }(450)$.
Matrix elements are evaluated in momentum space as in Refs.~\cite{Hagen2014,Arthuis2023}.

The normal-ordering scheme~\cite{Hagen2014,LietzCompNucl,Hebeler3nf} is an efficient way to incorporate part of the effects of 3N forces into an effective two-body interaction, $\Tilde{V}$, which has antisymmetrized matrix elements
\begin{align}
    \label{eq: eff 2N pot}
    \Tilde{v}_{\alpha\beta, \gamma\delta} = \Bar{v}_{\alpha\beta, \gamma\delta}
    + \sum_{h} \Bar{w}_{\alpha\beta h, \gamma\delta h},
\end{align}
where $h$ denotes the hole states.
The normal-ordered two-body (NO2B) approximation, employed in each many-body method, neglects the residual 3N interaction and is instrumental in reducing the cost of $\textit{ab initio}$ calculations, while introducing only a modest error~\cite{Djarv:2021xjg,Hagen2007,Hebeler3nf}.

\subsection{Coupled-cluster}
\label{sec: CC}

The coupled-cluster method~\cite{Hagen2014,Hagen2014Review,ShavittBartlett} postulates an exponential ansatz for the g.s. of many-nucleon systems:
\begin{align}
    \ket{\Psi} = e^{T} \ket{\Phi_0},
\end{align}
where the cluster operator $T$ is a superposition of $n$-particle-$n$-hole excitations on top of a specific reference state $\ket{\Phi_0}$, often taken to be Hartree-Fock (HF) solutions.

\begin{figure}[h!]
    \centering
    \includegraphics[width=\columnwidth]{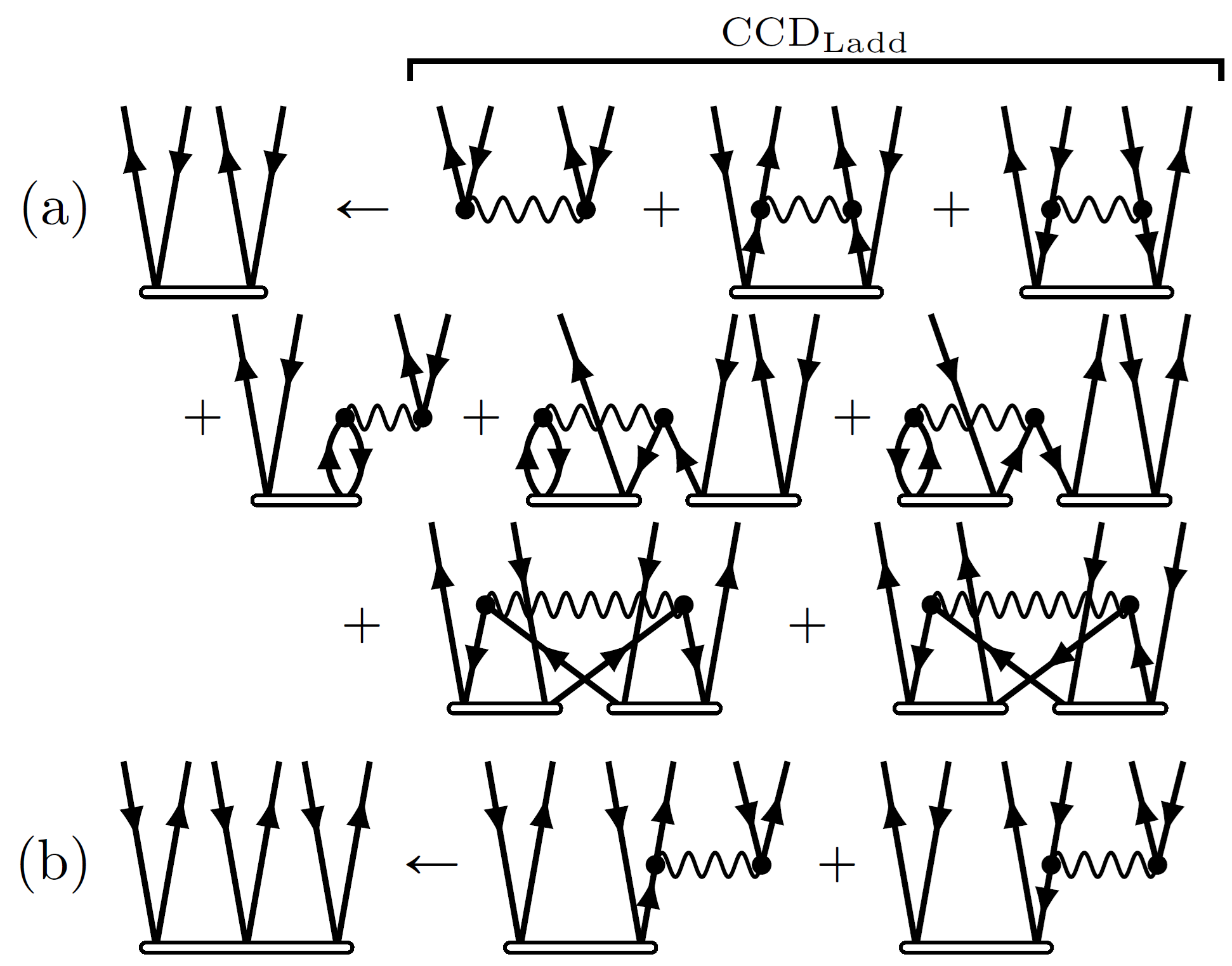}
    \caption{
        (a) Diagrammatic representation of the CCD equations for the doubles amplitude $T_2$.
        Wiggly lines denote the normal-ordered two-body interaction~\eqref{eq: eff 2N pot}, and amplitudes are shown as double horizontal lines. The CCD$_{\rm Ladd}$ approximation is also shown. See text for details. \\
        (b) Diagrams for the triples amplitude $T_3$ in the CCD(T) approximation, Eq.~\eqref{eq: t 3p3h no2b}. $T_3$ is expressed in terms of the converged CCD $T_2$ amplitudes.
    }
    \label{fig: CCD equations}
\end{figure}

The simplest truncation for $T$ includes $2p2h$ excitations only, 
\begin{align}
    T \approx T_2 = \frac{1}{4} \sum_{ \substack{ ab \\ ij} } t^{ab}_{ij} c_a^\dagger c_b^\dagger c_j c_i,
\end{align}
where $a,b$ ($i,j$) denote particle (hole) s.p. states, and defines the so-called CC at doubles level (CCD) scheme.
Note that in infinite matter, $1p1h$ excitations do not contribute due to momentum conservation in the plane-wave basis.
The unknown amplitudes are found as the solution to a set of non-linear equations, $\bra{\Phi_{ij}^{ab}} e^{-T} H_N e^{T} \ket{\Phi_0}=0$, which are represented in diagrammatic form (refer to~\cite{ShavittBartlett} for the interpretation rules and~\cite{Hagen2014Review,Bartlett2007} for the corresponding expressions) in Fig.~\ref{fig: CCD equations}.
The equations are represented in the form $T_2 = f(T_2)$, see~\cite{Baardsen2013,LietzCompNucl}, and are solved iteratively.
Note that terms involving insertions of the 1B Fock matrix are not shown, since in infinite matter they only contribute to the energy denominator. Also shown in Fig.~\ref{fig: CCD equations} is the more simple ladder approximation, CCD$_{\rm Ladd}$~\cite{Baardsen2013,LietzCompNucl}, which consists of just the $pp$ and $hh$ linear diagrams.

The CCD correlation energy is given as a function of the converged $T_2$ amplitudes by
\begin{align}
    \label{eq: eccd}
    \Delta E_{CCD} = \frac{1}{4} \sum_{ \substack{ ab \\ ij} } t^{ab}_{ij} \Tilde{v}_{ij,ab}.
\end{align}
As in ~\cite{Hagen2014}, CCD equations are solved at the level of NO2b forces, i.e. explicit 3N interactions enter the Fock matrix, the effective NN force and the reference energy, but are neglected from the CC diagrams.

A full inclusion of triples ($3p3h$) excitations is rather demanding, and their effect is instead included perturbatively in the CCD(T) approximation~\cite{Hagen2014,CrawfordSchaefer2007,ShavittBartlett}, which adds a (negative) correction to the CCD energy given by
\begin{align}
    \label{eq: E triples}
    \Delta E_{T} =
    \frac{1}{36} \sum_{ \substack{ abc \\ ijk} } 
    \epsilon^{ijk}_{abc} 
    \abs{t_{ijk}^{abc} }^2,
\end{align}
with $\epsilon^{ijk}_{abc} = \epsilon_{i} + \epsilon_{j} + \epsilon_{k} - \epsilon_{a} - \epsilon_{b} - \epsilon_{c}$ and $\epsilon$ denoting the HF s.p. energies.
The $t_{ijk}^{abc}$ amplitudes are approximated as a function of the $T_2$ amplitudes as
\begin{align}
    \label{eq: t 3p3h no2b}
    t_{ijk}^{abc} \approx \frac{1}{\epsilon^{ijk}_{abc}} \mel{ \Phi_{ijk}^{abc} }{ \Tilde{V} T_2 }{ \Phi_0 },
\end{align}
which is shown diagrammatically in Fig.\ref{fig: CCD equations}(b).
With two-body interactions, CCD(T) is complete up to fourth order in the MBPT~\cite{CrawfordSchaefer2007,ShavittBartlett}, and is considered a "gold standard" in \textit{ab initio} calculations of g.s. energies.

%\onecolumngrid
%\twocolumngrid

\subsection{Self-consistent Green's functions}
\label{sec: scgf}

The SCGF method aims at solving the Dyson or Gorkov equations for the one-body (1B) propagator or Green's function (GF) $g(\omega)$, which gives access to the total g.s. energy, the g.s. expectation values of all 1B observables, e.g. the momentum distribution, as well as properties of the neighboring $A \pm 1$ systems~\cite{Barbieri2004,Barbieri2017,Soma2020,Rios2020}.
While these equations are formally exact, they contain the self-energy operator $\Sigma^{\star}$, whose expression must be approximated in practical calculations by including appropriate classes of Feynman diagrams.
$\Sigma^{\star}$ depends on the frequency and two s.p. states (either holes or particles), $\Sigma^{\star} = \Sigma^{\star}_{\alpha\beta}(\omega)$.
In SCGF, the self-energy is expanded in terms of the dressed propagator itself, $\Sigma^{\star}=\Sigma^{\star}[g](\omega)$, which simultaneously determines and is determined by $\Sigma^{\star}$, and thus the method is inherently self-consistent. 
With this choice, the self-energy expansion is restricted to the so-called "skeleton" diagrams~\cite{Barbieri2017}.

In this work, we make use of the truncation scheme for Gorkov Green's functions developed in Refs.~\cite{Marino2024, MarinoPhdThesis}, which is based on the following equations:
\begin{align}
    \label{eq: Gorkov energy dependent}
    & 
    \begin{pmatrix}
        T - \mu \IdentityMat + \Sigma^{11}(\omega) & \Sigma^{12(\infty)} \\
        ( \Sigma^{12(\infty)})^{\dagger} & - (T - \mu \IdentityMat ) + \Sigma^{22}(\omega)
    \end{pmatrix}
    \eval_{\omega = \omega_q}
    \begin{pmatrix}
        \mathcal{U}^{q} \\ \mathcal{V}^{q}
    \end{pmatrix} 
    \nonumber \\
    & = \hbar \omega_q
    \begin{pmatrix}
        \mathcal{U}^{q} \\ \mathcal{V}^{q}
    \end{pmatrix} .
\end{align}

The eigenvalues $\omega_q$ and amplitudes $\mathcal{U}^{q}$, $\mathcal{V}^{q}$ completely determine the propagator.
The chemical potential $\mu$ is adjusted to reproduce on average the correct number of particles for each fermion species.
In our specific implementation, pairing correlations are handled at first order, i.e. the off-diagonal anomalous self-energy is approximated by the frequency-independent function $\Sigma^{12(\infty)}$.
The bulk of many-body correlations is encoded in the normal self-energy $\Sigma^{11}(\omega)$, which can be decomposed into the sum of the static contribution $\Sigma^{11(\infty)}$, that can be interpreted as a generalized mean field for the interacting system, and a frequency-dependent term, $ \Tilde{\Sigma}^{11}(\omega)$.
The dynamical self-energy is approximated using the state-of-the-art algebraic diagrammatic construction (ADC) scheme at third order on top of a HF-like mean-field state [Dyson-ADC(3)]~\cite{Soma2020,Raimondi2017}.
This hybrid scheme is an effective way of including pairing effects and dynamical correlations at the same time, while being more manageable than the full Gorkov-ADC(3) method~\cite{Barbieri2022Gorkov}.

\begin{figure}[ht!]
    \centering
    \includegraphics[width=\columnwidth]{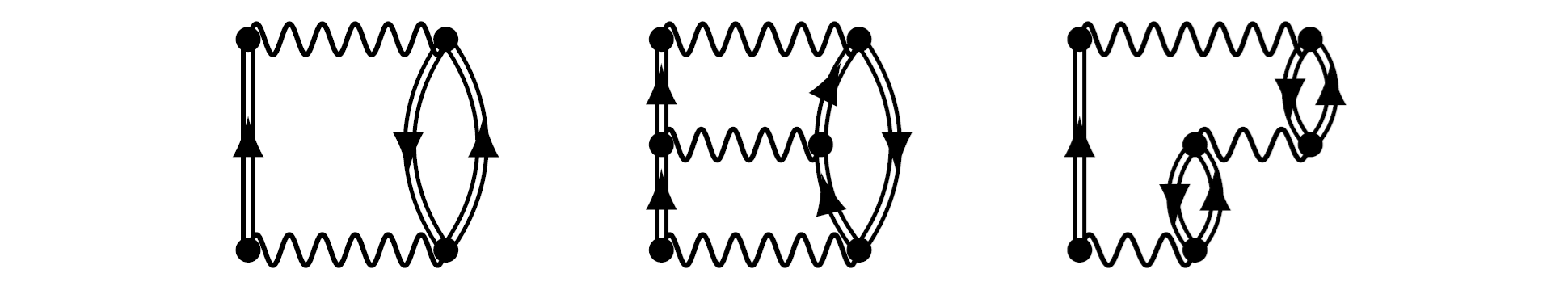}
    \caption{
    Diagrams defining the dynamical self-energy $\Tilde{\Sigma}(\omega)$ in the Dyson-ADC(3) approximation.
    From left to right, the second-order, the third-order ladder, and the third-order ring diagrams are shown.
    Wiggly lines and double lines represent NO2B interactions and dressed propagators, respectively. 
    ADC(2) includes just the second-order term.
    The ADC(ld,2) truncation is defined by including also the ladder diagram but neglecting the ring (Sec.~\ref{sec: truncations}).
    Rules for evaluating the Feynman diagrams are discussed in e.g. Refs.~\cite{Barbieri2017,Raimondi2017}. 
    }
    \label{fig: ADC(3) diags}
\end{figure}

ADC~\cite{Schirmer2018,Raimondi2017,Barbieri2017} defines a systematically improvable hierarchy of approximations for the self-energy, ADC($n$), which by construction respect the analytical structure of the exact $\Sigma^{\star}$ and are consistent with perturbation theory up to $n$-th order.
In particular, ADC(3) incorporates all third-order self-energy diagrams featuring only (effective) two-body interactions, and at the same time resums infinite classes of diagrams automatically, such as the ladders and (Tamm-Dancoff) rings, plus specific fourth-order diagrams. 
The 'seed' diagrams for the ADC(3) dynamical self-energy are shown in Fig.~\ref{fig: ADC(3) diags}.

The static self-energies $\Sigma^{11(\infty)}$ and $\Sigma^{12(\infty)}$ are computed exactly in terms of the bare NN and 3N interactions and the dressed propagator, and they are iterated fully.
The dynamical self-energy, instead, is evaluated at the NO2B level, neglecting irreducible 3N contributions. 
With these assumptions, the ADC(3) matrices propagate two-particle-one-hole ($2p1h$) and $2h1p$ intermediate configurations.
Moreover, in contrast with the static part, $\Tilde{\Sigma}^{11}(\omega)$ is not built out of the complete $g(\omega)$, but rather out of an uncorrelated GF, dubbed the optimized reference state (OpRS), that is designed to accurately reproduce the dressed propagator at a reduced computational cost~\cite{Barbieri2009,Barbieri2022Gorkov}.
In practice, an iterative solution to Eq.~\eqref{eq: Gorkov energy dependent} is sought as follows.
$\Tilde{\Sigma}^{11}(\omega)$ is evaluated once and stored; this is the most expensive part of the method.
Then, the static self-energies are updated repeatedly and a diagonalization is performed at each iteration ("sc0" loop). When the total energy has converged within a given tolerance, the energies of s.p. states are readjusted, the dynamical self-energy is evaluated again using the new reference, and a new sc0 cycle is performed.
The prescription we use to determine the optimized s.p. energies (called "Cen-PH" in~\cite{Marino2024}) requires first evaluating the centroid energy for each state $\alpha$ ($\Bar{\omega}_\alpha$) by weighting the poles $\omega_q$ by the spectral distribution, proportional to the imaginary part of $g(\omega)$.
Next, we assign energies $\epsilon_\alpha = \mu \pm \Bar{\omega}_\alpha$ to states that are unoccupied (occupied) in the HF g.s.

An important extension of ADC(3), which combines SCGF with CC, is ADC(3)-D~\cite{Soma2013,Barbieri2017,Marino2024,Marino2024Qnp}. The ADC(3) dynamical self-energy evaluated on a mean-field state (see e.g.~\cite{Raimondi2017,Marino2024}) depends on the contraction between interaction matrix elements and unperturbed $2p2h$ amplitudes $\kappa$, defined in Eq.~\eqref{eq: amplitude mbpt}, which are also the zeroth-order approximation for the CC $T_2$ amplitudes.
This suggests that a wider set of correlations can be included by a modification of ADC(3) in which $\kappa$ is replaced with the converged $t$ amplitudes obtained as a solution of the CCD equations~\cite{Barbieri2017,Hodecker2019}. 
The $t$'s generate additional infinite classes of diagrams which allow to improve the accuracy of the method in strongly-interacting regimes~\cite{Barbieri2017}, at the cost of performing a preliminary CCD computation.
In the following, results obtained with both ADC(3) and ADC(3)-D are presented.

\subsection{Many-body perturbation theory}
\label{sec: MBPT}

MBPT defines a hierarchy of order-by-order approximations to the correlation energy~\cite{ShavittBartlett,Drischler2021Review,Tichai2020Mbpt}.
With soft chiral interactions, MBPT (in the thermodynamic limit formulation) has proven a useful tool to investigate nuclear matter, see e.g.~\cite{Drischler2019,Keller2023,Drischler2021Review}, and a reasonable convergence of the perturbative expansion may be achieved already at third or, at most, fourth order~\cite{Drischler2019}.
Therefore, the relatively simple MBPT(3) method constitutes a computationally cheaper alternative to more sophisticated non-perturbative techniques, such as SCGF, CC, or CIMC. Neutron matter, in particular, can be successfully tackled in this way. 

\begin{figure}[h!]
    \centering
    \includegraphics[width=\columnwidth]{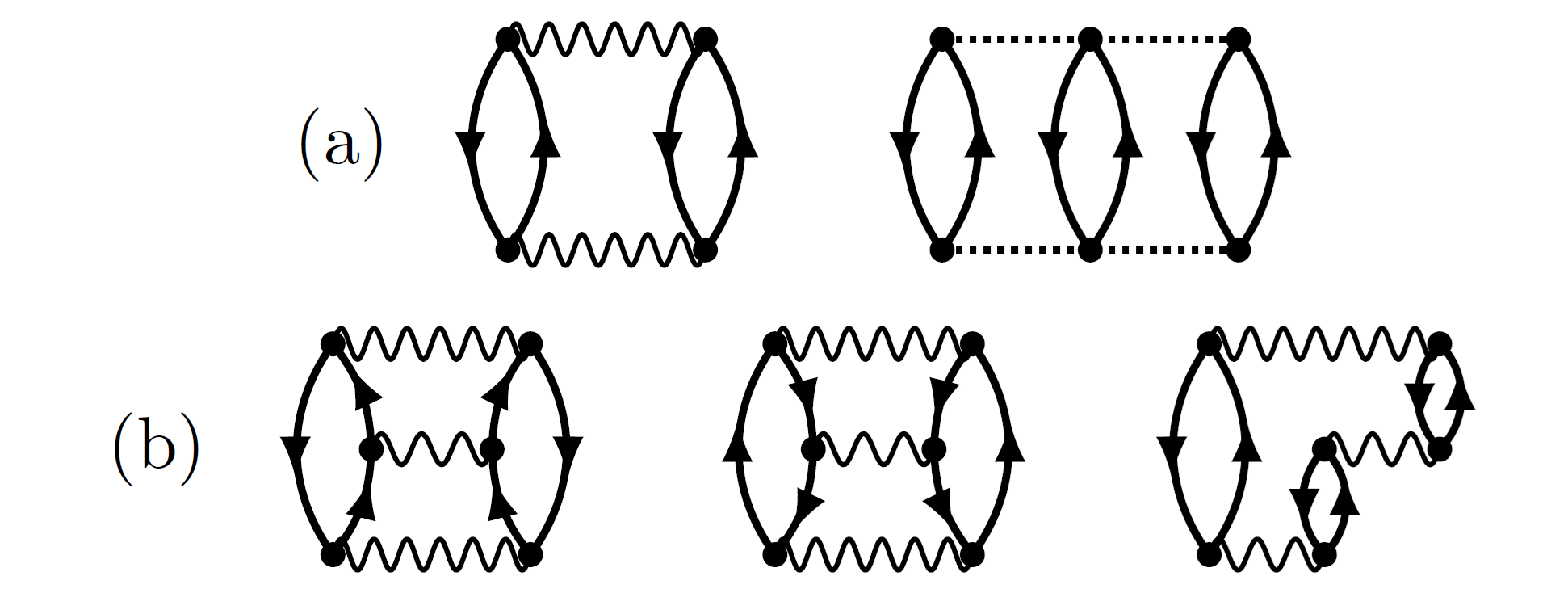}
    \caption{
    Diagrammatic representation of MBPT contributions to the correlation energy up to the third order.
    In the top row (a), the second-order NN diagram~\eqref{eq: mbpt2 NN} and the irreducible 3N term~\eqref{eq: mbpt2 3N} are shown to the left and right, respectively.
    In the bottom row (b), third-order contributions are shown. From left to right: $pp$~\eqref{eq: mbpt3 pp}, $hh$~\eqref{eq: mbpt3 hh}, and $ph$~\eqref{eq: mbpt3 ph} ladder terms. 
    Wiggly lines denote the normal-ordered two-body interaction, while dashed lines refer to bare 3N potentials.
    Rules for interpreting the diagrams can be found e.g. in Refs.~\cite{ShavittBartlett,Drischler2021Review}.
    }
    \label{fig: mbpt diagrams}
\end{figure}

At second order, NN and 3N forces are included in full. For the third-order terms, the NO2B approximation is adopted, i.e. we evaluate the three ladder diagrams ($pp$, $hh$, $ph$) that include only two-body interactions and neglect all those terms that feature irreducible 3N insertions.
This approximation, which is equivalent to the one used in Ref.~\cite{Keller2023}, is represented diagrammatically in Fig.~\ref{fig: mbpt diagrams} and can be written as follows:
\begin{align}
    \Delta E^{(3)} = \Delta E^{(2,NN)} + \Delta E^{(2,3N)} + \Delta E^{(3,NN)},
\end{align}
where $\Delta E^{(2,NN)}$ is the standard second-order term,
\begin{align}
    \label{eq: mbpt2 NN}
    \Delta E^{(2,NN)} = \frac{1}{4}
    \sum_{abij} \frac{ \abs{ \Tilde{v}_{ab,ij} }^2 }{ \epsilon^{ij}_{ab} } =
    \frac{1}{4}
    \sum_{abij} \kappa^{ab}_{ij} \Tilde{v}_{ij,ab},
\end{align}
and $\Delta E^{(2,3N)}$ is the irreducible 3N second-order diagram~\cite{Hebeler2010},
\begin{align}
    \label{eq: mbpt2 3N}
    \Delta E^{(2,3N)} = \frac{1}{36}
    \sum_{ \substack{ abc \\ ijk} } \frac{ \abs {\Bar{w}_{abc,ijk} }^2 }{ \epsilon^{ijk}_{abc} }.
\end{align}
There are three diagrams that contribute at third order\cite{ShavittBartlett,LietzCompNucl},
%and the three diagrams of Fig.~\ref{fig: mbpt diagrams}(b) contribute at third order~\cite{ShavittBartlett,LietzCompNucl},
\begin{align}
    \Delta E^{(3,NN)} = \Delta E^{(3,pp)} + \Delta E^{(3,hh)} + \Delta E^{(3,ph)},
\end{align}
which are shown in Fig.~\ref{fig: mbpt diagrams}(b) and can be explicitly computed from
\begin{align}
    \label{eq: mbpt3 pp}
    & \Delta E^{(3,pp)} = \frac{1}{8} \sum_{ \substack{abcd \\ ij } } 
    ( \kappa^{ab}_{ij} )^{*} \Tilde{v}_{ab,cd} \kappa^{cd}_{ij},
    \\
    \label{eq: mbpt3 hh}
    & \Delta E^{(3,hh)} = \frac{1}{8} \sum_{ \substack{ab \\ ijkl} } \kappa_{ij}^{ab} \Tilde{v}_{ij,kl} (\kappa^{ab}_{kl})^{*}, \\
    \label{eq: mbpt3 ph}
    & \Delta E^{(3,ph)} = \quad \sum_{ \substack{abc\\ ijk} }
    ( \kappa^{ab}_{ij} )^{*} \Tilde{v}_{bk,jc} \kappa^{ac}_{ik}.
\end{align}
A compact expression for the above equations has been achieved by introducing the amplitude 
\begin{align}
    \label{eq: amplitude mbpt}
    \kappa^{ab}_{ij} = \frac{ \Tilde{v}_{ab,ij} }{ \epsilon^{ij}_{ab}},
\end{align}
with $\epsilon^{ij}_{ab} = \epsilon_i + \epsilon_j - \epsilon_a - \epsilon_b$.
Note that we use HF s.p. energies for the MBPT energy denominators, and Eq.~\eqref{eq: amplitude mbpt} is the initial approximation for the CC amplitudes.

\section{Results}
\label{sec: results}

In this section, benchmark calculations are presented for a selection of realistic chiral interactions at NNLO, all of which include both NN and 3N contributions (Sec.~\ref{sec: methods}). 
Results are reported for the CC method with both the CCD and CCD(T) approximations (Sec.~\ref{sec: CC}), for ADC-SCGF using the ADC(3) truncation of the self-energy in terms of skeleton diagrams and the ADC(3)-D scheme (Sec.~\ref{sec: scgf}), and for MBPT at third order (Sec.~\ref{sec: MBPT}).
Matter densities $\rho$ between 0.04 and 0.28 $\rm{fm}^{-3}$ are considered; this range is sufficiently large to allow for a methodological comparison also in regimes of strong correlations and lies within the region of validity for chiral effective field theories.

We emphasize that the results presented in the following sections reflect only the errors arising from the many-body methods and the possible cutoff dependence of $\chi$EFT.
Uncertainties due to the model discrepancies~\cite{PbAbInitio,Jiang2024} and the order-by-order expansion of $\chi$EFT~\cite{Drischler2020,Acharya:2022drl,Tews:2024owl} are not included, as these are beyond the scope of this benchmark study. For detailed predictions, particularly those concerning saturation properties, these factors are currently expected to be the predominant sources of residual error.

\subsection{Pure neutron matter}
\label{sec: pnm results}

\begin{figure}[h!]
    \centering
    \includegraphics[width=\columnwidth]{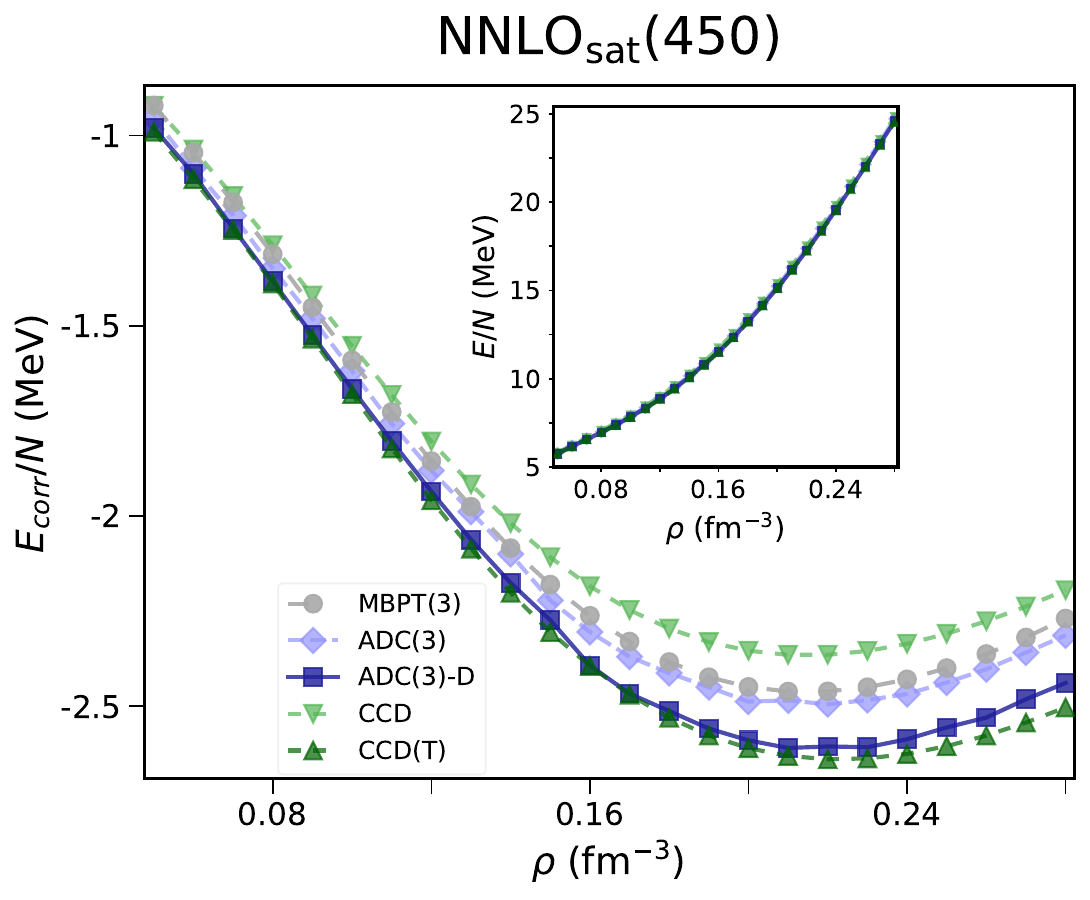}
    \caption{
    Correlation energy per particle (main panel) and energy per particle (inset) as a function of the number density in PNM. 
    Calculations are performed with several many-body methods (see legend) using the $\rm{ NNLO_{sat} }(450)$ interaction with $N=66$ neutrons subject to PBCs.
    }
    \label{fig: nnlosat PNM inset}
\end{figure}

We start with a study of the PNM EOS using the $\rm{ NNLO_{sat} }(450)$ interaction~\cite{NNLOsat} (Fig.~\ref{fig: nnlosat PNM inset}). 
This is a relatively hard potential, making it conducive for emphasizing the differences between techniques.
However, the weakly correlated nature of neutron matter is apparent from the total energies (shown in the inset), where the different curves closely overlap at all densities. Differences are better highlighted from the correlation energies, $E_{corr}=E-E_{HF}$, which are shown in the main panel.
The highest and lowest energies are those obtained with CCD (upward arrows) and CCD(T) (downward arrows), respectively.
MBPT(3) (circles) and ADC(3) (diamonds) lie at intermediate values within the band spanned by the CC approximations and are very close to each other.
Interestingly, the inclusion of CC amplitudes in ADC(3)-D (squares) lowers the ADC(3) energies and brings the curve almost on top of CCD(T). 
Correlation energies tend to increase in magnitude from -1 $\rm{MeV}/N$ at $\rho=0.04\,\rm{fm}^{-3}$ to about -2.5 $\rm{MeV}/N$ at $\rho=0.20\,\rm{fm}^{-3}$.
In general, there is a very good agreement between all the techniques, and differences are of the order of a few hundred keV at most at saturation density and even above.
This comparison suggests that PNM is well under control from the perspective of the many-body methods, and uncertainties stem predominantly from the nuclear interaction.

The predictions for the PNM EOS obtained with a selection of chiral forces are compared in Fig.~\ref{fig: many pnm eos}.
The outcomes for different methods and interactions are also reported in the Supplemental Material.
Results are shown as bands that encompass the range between the results obtained with CCD(T) and ADC(3).
These bands provide an estimate of the precision attainable in solving the \Sch equation in PNM and are very thin for all the nuclear potentials considered, as was already noted for the case of $\rm{ NNLO_{sat} }$. In contrast, it is known that the $\rm{ NNLO_{sat} }$ PNM EOS is too soft and the symmetry energy too small~\cite{Carbone2020}, while the delta-full models mitigate the issue~\cite{DeltaGo2020} by providing more realistic values of the symmetry energy at saturation and look more consistent with each other in this respect.
An additional outcome of this analysis is that MBPT(3) in a finite model space is a valid and cheaper alternative to CC and SCGF for neutron matter.

\begin{figure}[h!]
    \centering
    \includegraphics[width=\columnwidth]{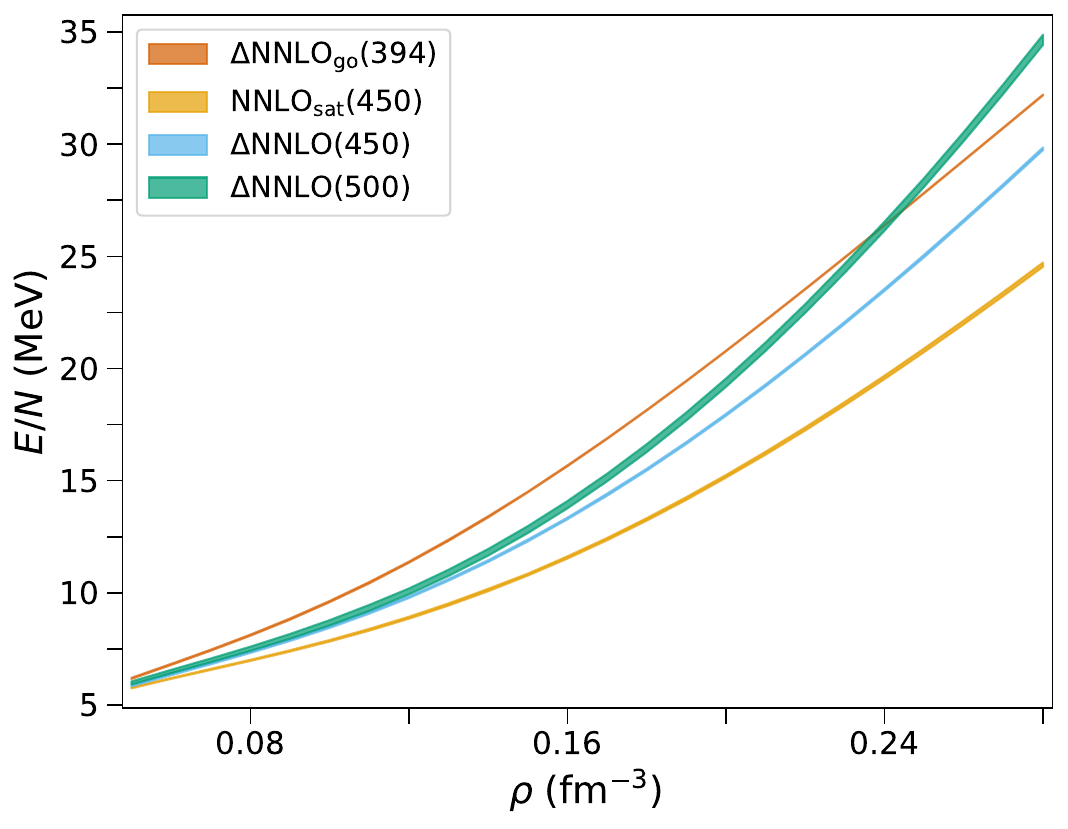}
    \caption{
    PNM equations of state for a selection of chiral interactions.
    Predictions are shown as shaded curves representing the interval between the predictions of CCD(T) and ADC(3).
    }
    \label{fig: many pnm eos}
\end{figure}

\subsection{Symmetric nuclear matter}
\label{sec: snm results}

\begin{figure*} %[h!]
    \centering
    \includegraphics[width=0.45\textwidth]{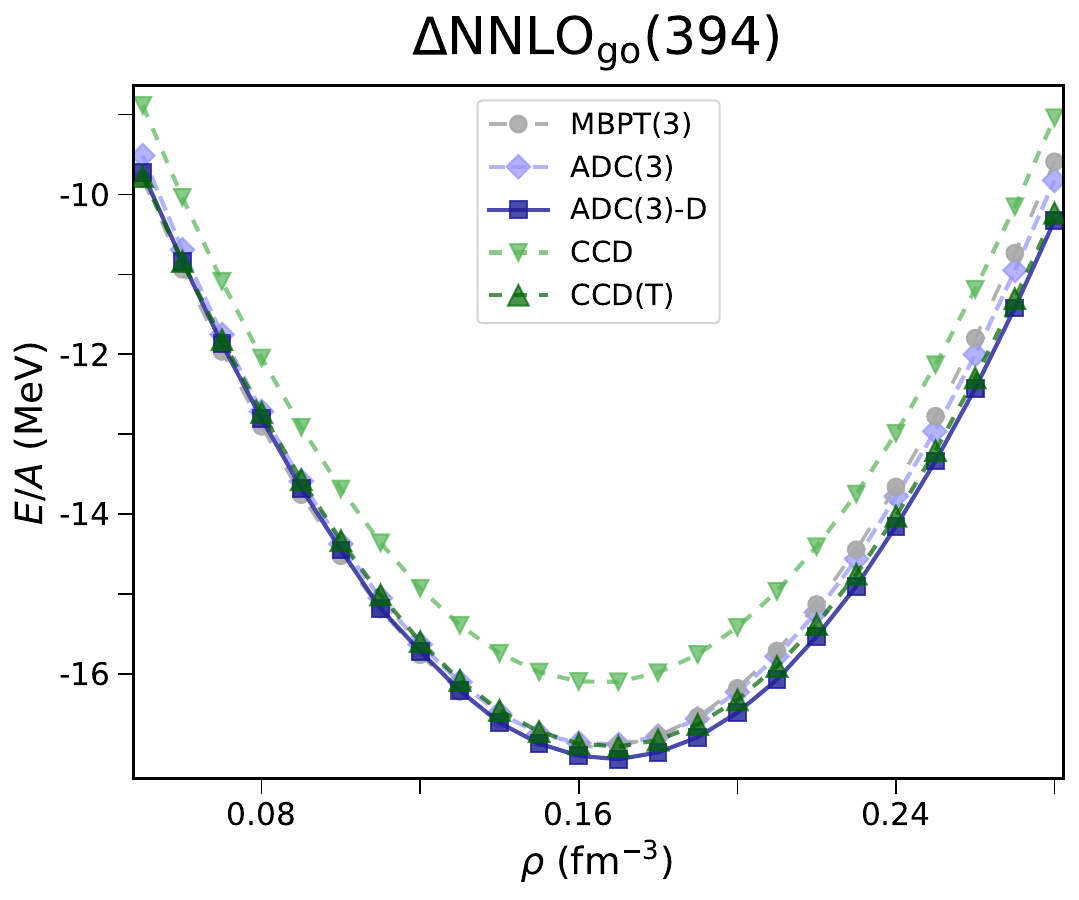}
    \includegraphics[width=0.45\textwidth]{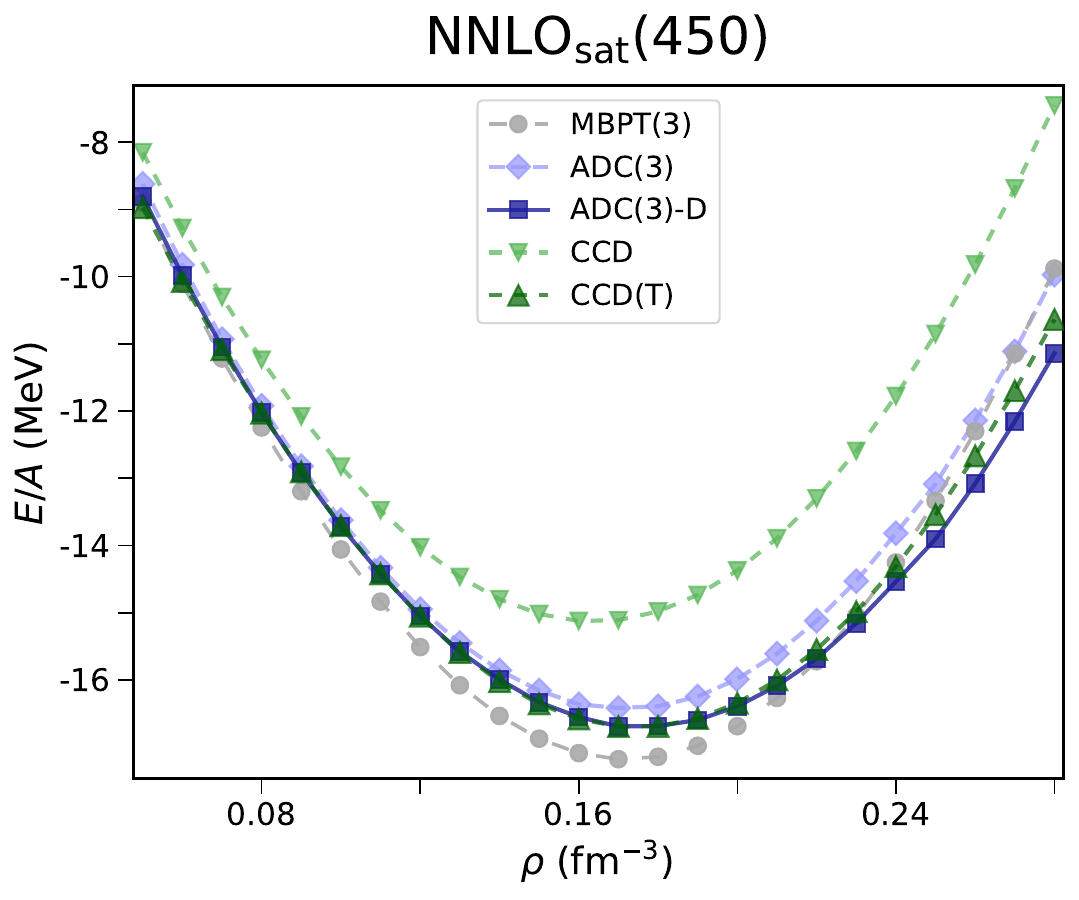}
    \includegraphics[width=0.45\textwidth]{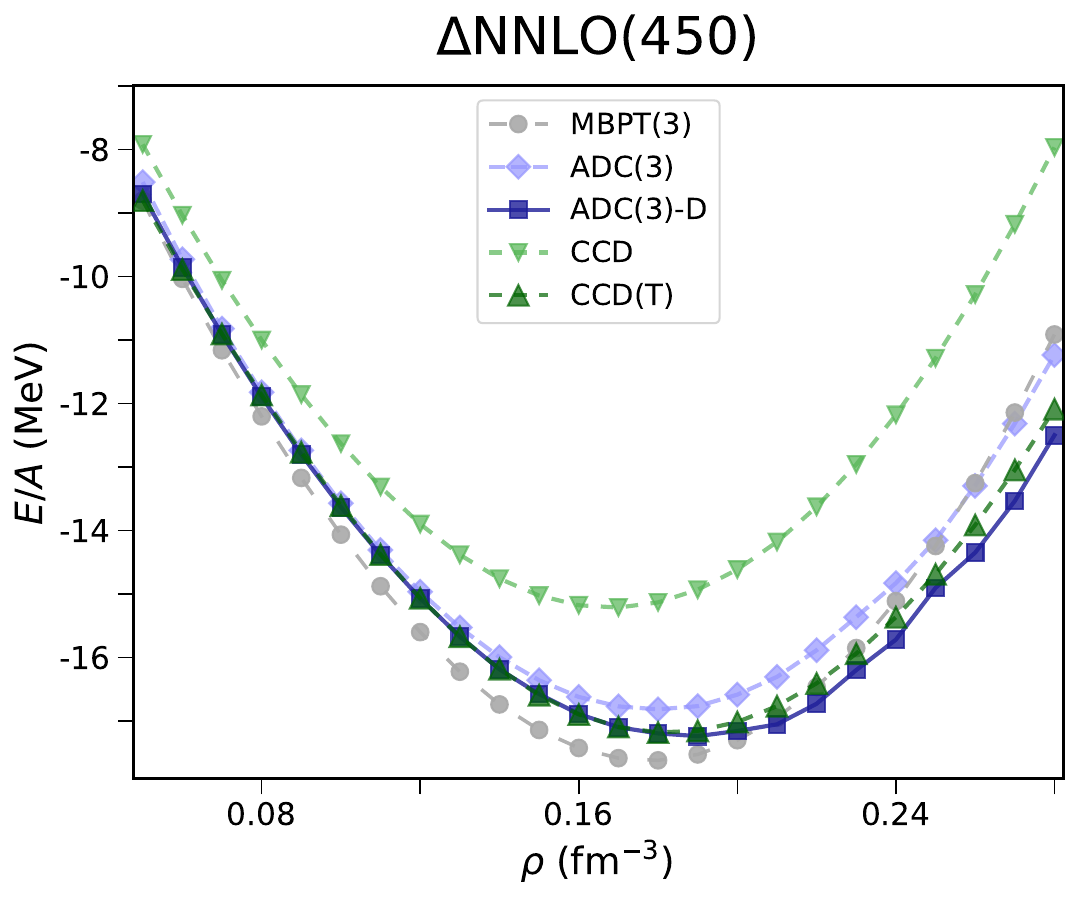}
    \includegraphics[width=0.45\textwidth]{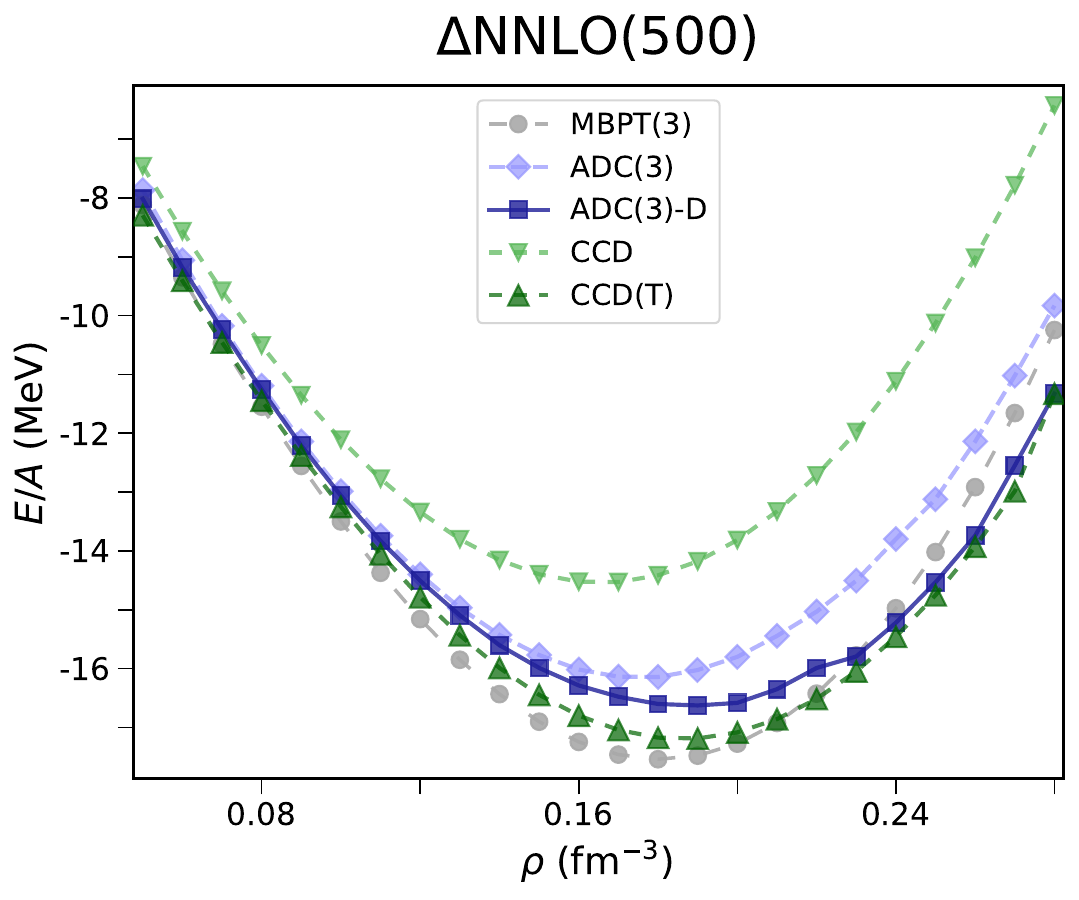}
    \caption{
    SNM equations of state obtained with four different chiral interactions.
    Each panel reports the outcomes of five many-body techniques (see legend).
    $A=132$ nucleons subject to PBCs are employed to simulate SNM.
    }
    \label{fig: SNM panels}
\end{figure*}

Symmetric matter is a strongly correlated system and represents a more demanding test for \textit{ab initio} methods.
While, for a given interaction, all techniques essentially converge to the same result in PNM, larger method differences are in general observed for SNM, although the magnitude of their discrepancy is somewhat dependent on the potential.
Each panel of Fig.~\ref{fig: SNM panels} reports the EOS for a different chiral interaction, with five curves denoting the outcomes of the different methods: MBPT(3), CCD, CCD(T), ADC(3), and ADC(3)-D.
These results are also reported in tabular form in the Supplemental Material.
The four potentials are ordered by increasing values of their cutoff, with $\rm{ \Delta NNLO_{go} (394)}$ (top left) being the softest interaction and $\rm{ \Delta NNLO (500)}$ (bottom right) being the hardest.
Note that, in any case, chiral forces are considerably softer than the phenomenological potentials, such as the Argonne models~\cite{Wiringa1995,Wiringa2002}, even without similarity-renormalization-group (SRG) evolution~\cite{Hergert2020}, as regularization suppresses the hard repulsive core.
This is essential to ensuring good convergence with respect to both the many-body truncation and the model space dimension for methods rooted in a basis expansion.
(Coordinate-space Monte Carlo methods such as AFDMC can handle the hard core relatively more easily, but studies with fully realistic interactions are somewhat limited to PNM, with few exceptions~\cite{Lonardoni2020LocalChiral}.
Finite-temperature SCGF, which is based on a continuum of momentum states, can also handle hard potentials~\cite{Rios2020,Rios2017}.)

Additionally, while it is generally known that particle-particle ladder diagrams--which are resummed in Br\"{u}cker theory, as well as in CC and SCGF--should be the dominant contributions to the nuclear matter energy~\cite{Mattuck,Baldo2012}, their relevance is somewhat reduced with low-momentum interactions~\cite{Drischler2021Review}, and finite-order perturbation theory becomes viable even for SNM. 

The case of $\rm{ \Delta NNLO_{go} (394)}$ is noteworthy because the energies per nucleon yielded by the two ADC variants, CCD(T), and MBPT(3) are rather close to each other.
In contrast, the CCD truncation is insufficient, and triples corrections are noticeable around saturation density, contributing about $-0.8\,\rm{MeV}$ to the energy per nucleon. 
As for PNM, ADC(3) is always on top of CCD(T) results, but differences remain small, of the order of 200 keV or less, while ADC(3)-D matches almost exactly with CCD(T). Notably, MBPT(3) compares very well with the other techniques.
Overall, SNM calculations for the $\rm{ \Delta NNLO_{go} (394)}$ interactions are all consistent with each other and provide convincing evidence that the three diagrammatic methods can all handle correlations accurately.
Of course, this interaction is very soft and particularly well-suited for perturbative techniques.
Note also that ADC-SCGF computations converge quickly in this case, typically within 5 OpRS cycles, and the optimized s.p. energies remain close to the HF ones. 

Next, results for the $\rm{ NNLO_{sat} }(450)$ (top right) and $\rm{ \Delta NNLO (450)}$ (bottom left) interactions are considered.
A consistent pattern is recognized at all densities, with CCD yielding the highest energies per particle and ADC(3) returning energies always higher than those from CCD(T).
In addition, the agreement between CCD(T) and ADC(3)-D is excellent. Discrepancies are noticeable only at high density, while the EOS is rather well constrained below $\rho_{0}$.
As for MBPT(3), on the one hand, perturbative results provide a qualitatively reasonable estimate of the EOS. 
On the other hand,  it is difficult to find a well-defined quantitative trend, as MBPT(3) is more attractive than CCD(T) up to approximately the saturation point but becomes more repulsive at higher densities.

The case of the $\rm{ \Delta NNLO (500)}$ is the most challenging, as the interaction is the hardest among those considered and induces strong many-particle correlations.
For instance, the triples correction~\eqref{eq: E triples} is as large as about -2.3 MeV at $\rho_0=0.16\,\rm{fm}^{-3}$, more than twice as much than for $\rm{ \Delta NNLO_{go} (394)}$, and increases to -3.3 MeV at $\rho=0.20\,\rm{fm}^{-3}$, in contrast to -0.9 MeV for $\rm{ \Delta NNLO_{go} (394)}$.
The magnitude of $\Delta E_{T}$ and its rapid growth as a function of $\rho$ suggest that contributions from $3p3h$ configurations are significant.
It is not unexpected that we observe significant discrepancies between methods.
As always, ADC(3) energies are intermediate between those from CCD and CCD(T), but differ from the latter by almost $1\,\rm{MeV}$ around saturation density.
ADC(3)-D also provides an upper bound to CCD(T) energies, though deviations are overall smaller than for ADC(3). 
At first sight, MBPT(3) results seem relatively close to CCD(T), yet this may be coincidental.
Indeed, MBPT(3) underestimates the CCD(T) saturation energy, which is found at a large density $> 0.18\,\rm{fm}^{-3}$ and low binding energy. However, it provides insufficient correlation energy in the high-density region.
We infer that the role of fourth-order perturbative corrections should be sizeable, and suggest that differences between CCD(T) and ADC could be due to each method including a different subset of these contributions.
As a further comment, though, we also point out that the magnitude of the triples correction puts into question the validity of the CCD(T) itself, which should be checked against iterative CC schemes comprising $3p3h$ amplitudes, such as CCDT-1~\cite{Hagen2014Review}.

In summary, $\rm{ \Delta NNLO (500)}$ is a challenging test for many-body approaches.
While tighter quantitative constraints can be found for softer interactions, systematic method uncertainties are considerable even around saturation density for this interaction. 
An extension to even harder interactions, such as local chiral interactions~\cite{Piarulli2020Delta,Gezerlis2013} or the EMN550 potential~\cite{Entem2017}, would further amplify these challenges. The non-perturbative nature of such interactions increases the computational demands significantly, often requiring larger model spaces and technical improvements to achieve reliable convergence. However, within the present frameworks, achieving stable and converged results may not be feasible (we already observed long iterations and occasional numerical instabilities in CC and ADC for $\rm{ \Delta NNLO (500)}$). 

\onecolumngrid

%\twocolumngrid

\newpage

\twocolumngrid

\subsection{Computational aspects}
\label{sec: scaling}

\textit{Ab initio} techniques demand significant computational resources and careful numerical implementation. 
We employed two distinct codes: one for ADC(3), ADC(3)-D, and MBPT(3), and another for CCD and CCD(T). Both codes are highly parallelized, utilizing hybrid MPI/OpenMP programming models. 
The basic workflow of a calculation begins with constructing the s.p. basis, followed by the evaluation and storage of the matrix elements of the normal-ordered Hamiltonian, Eq.~\eqref{eq: eff 2N pot}. 
It is worth noting that we find it more efficient to compute the matrix elements at each run rather than loading them from a file, thanks to the use of effcient routines that evaluate the potentials directly in the momentum basis, thus avoiding the costly partial-wave expansions and the overhead of slow input/output operations.

Afterward, the actual many-body calculations commence.
In Tab.~\ref{tab: theor scaling}, we present theoretical estimates for the computational scaling of ADC, CC, and MBPT as a function of the number of hole states ($N_h$) and particle states ($N_p$) in the s.p. basis.
(In our case, $N_h = A$.)
ADC(3), CCD, and MBPT(3) formally exhibit the same scaling behavior. 
In ADC(3), the dominant factor in the computational workload is the matrix-vector product required in the Lanczos reduction of the dynamical self-energy in the $2p1h$ sector, see e.g.~\cite{Weikert,Banerjee2023,Schirmer2018}.
In MBPT(3), the most computationally demanding step is the evaluation of the $pp$ ladder diagram~\eqref{eq: mbpt3 pp}, which involves summing over 4 particle and 2 hole indices.
Also challenging is the irreducible 3N second-order correction, Eq.~\eqref{eq: mbpt2 3N}.
Similarly, in CCD calculations, the contribution to the amplitude equations from $pp-pp$ interactions represents the most expensive part of the process~\cite{LietzCompNucl,Hagen2014,ShavittBartlett}.
ADC(3)-D scales as the standard ADC(3), but of course it incurs an additional overhead due to the need to solve for the CCD ground state first. 
Finally, the non-iterative CCD(T) correction has a computational cost proportional to $N_h^3 N_p^4$, which arises from summing over $3p3h$ states in Eq.~\eqref{eq: E triples}, along with an extra summation over one particle index to generate the approximate triples amplitudes, see Fig.~\ref{fig: CCD equations}(b).
\noindent
\begin{minipage}{\columnwidth}
    \centering
    \begin{tabular}{ p{3cm} p{3cm} p{3cm} }
        Method & Cost & Refs. \\  
        \noalign{\vskip 1.mm} 
        \hline
        \noalign{\vskip 1.5mm} 
        ADC(3) & $N_h^2 N_p^4$ & \cite{Banerjee2023,Schirmer2018} \\
        \noalign{\vskip 1.5mm} 
        ADC(3)-D & $N_h^2 N_p^4$ &  \\
        \noalign{\vskip 1.mm} 
        CCD & $N_h^2 N_p^4$ & \cite{Hagen2014,LietzCompNucl} \\
        \noalign{\vskip 1.mm} 
        CCD(T) & $N_h^3 N_p^4$ & \cite{Bartlett2007} \\
        \noalign{\vskip 1.mm} 
        MBPT(3) & $N_h^2 N_p^4$ & \cite{ShavittBartlett}
    \end{tabular}
    \captionof{table}{Theoretical scaling of the computational cost as a function of the number of hole ($N_h$) and particle ($N_p$) states in the s.p. model space.
    References to the literature are reported in the third column.
    }
     \label{tab: theor scaling}
\end{minipage}

In practice, several key caveats that may influence the actual computational cost should be taken into account. First, the estimates shown in Tab.~\ref{tab: theor scaling} might be improved by considering momentum conservation~\cite{Hagen2014}, which is exploited in our codes and significantly reduces computational expense, similarly to using the angular-momentum-coupled (J-scheme) representation in finite nuclei~\cite{Hagen2014Review}.
Second, these estimates overlook the fact that setting up the NO2B Hamiltonian is a time-consuming task, often contributing a large portion of the total CPU time. 
Furthermore, achieving ideal efficiency can be challenging. For instance, while the construction of $3p3h$ states is parallelized, its scaling is hindered by communication inefficiencies between processes. As a consequence, evaluating the non-perturbative triples corrections~\eqref{eq: E triples} and ~\eqref{eq: mbpt2 3N} tends to be somewhat slower than anticipated.

We focus on the case of SNM at $\rho=0.16\,\rm{fm}^{-3}$ with $A=132$ and the  $\rm{ \Delta NNLO }(500)$ interaction, and discuss the actual run times in terms of total core hours (CPUh).
These calculations have been performed on the CINECA Galileo100 supercomputer utilizing 96 cores. 
Among the methods, CCD was the fastest, with a cost of about 40 CPUh, followed by MBPT(3) at 85 CPUh. 
Notably, ADC(3) turns out to be relatively efficient, with a runtime of 100 CPUh.
ADC(3)-D and CCD(T), on the other hand,  required 162 and 170 CPUh, respectively.

The impact of $3p3h$ diagrams is significant, as anticipated, contributing around 50 CPUh for MBPT(3) and 103 for CCD(T), accounting for more than 50\% of the total computational time in both cases.
In terms of both accuracy and computational cost, ADC(3)-D is comparable to CCD(T).
We note also that ADC(3) and ADC(3)-D computations both converged within a tolerance of less than 10 keV on the energy per particle in 3 OpRS cycles, each consisting of at most 8 sc0 loops (Sec.~\ref{sec: scgf}).
However, the required number of OpRS iterations can vary depending on the specific case, though in our experience 10 cycles often suffice to reach this level of numerical accuracy.
In contrast, CCD is much computationally lighter, and an accuracy of the order of $10^{-8}$ on the total energy is obtained in roughly 20 iterations of the amplitude equations.

\subsection{Comparison of many-body truncations}
\label{sec: truncations}

\begin{figure}[ht!]
    \centering
    \includegraphics[width=\columnwidth]{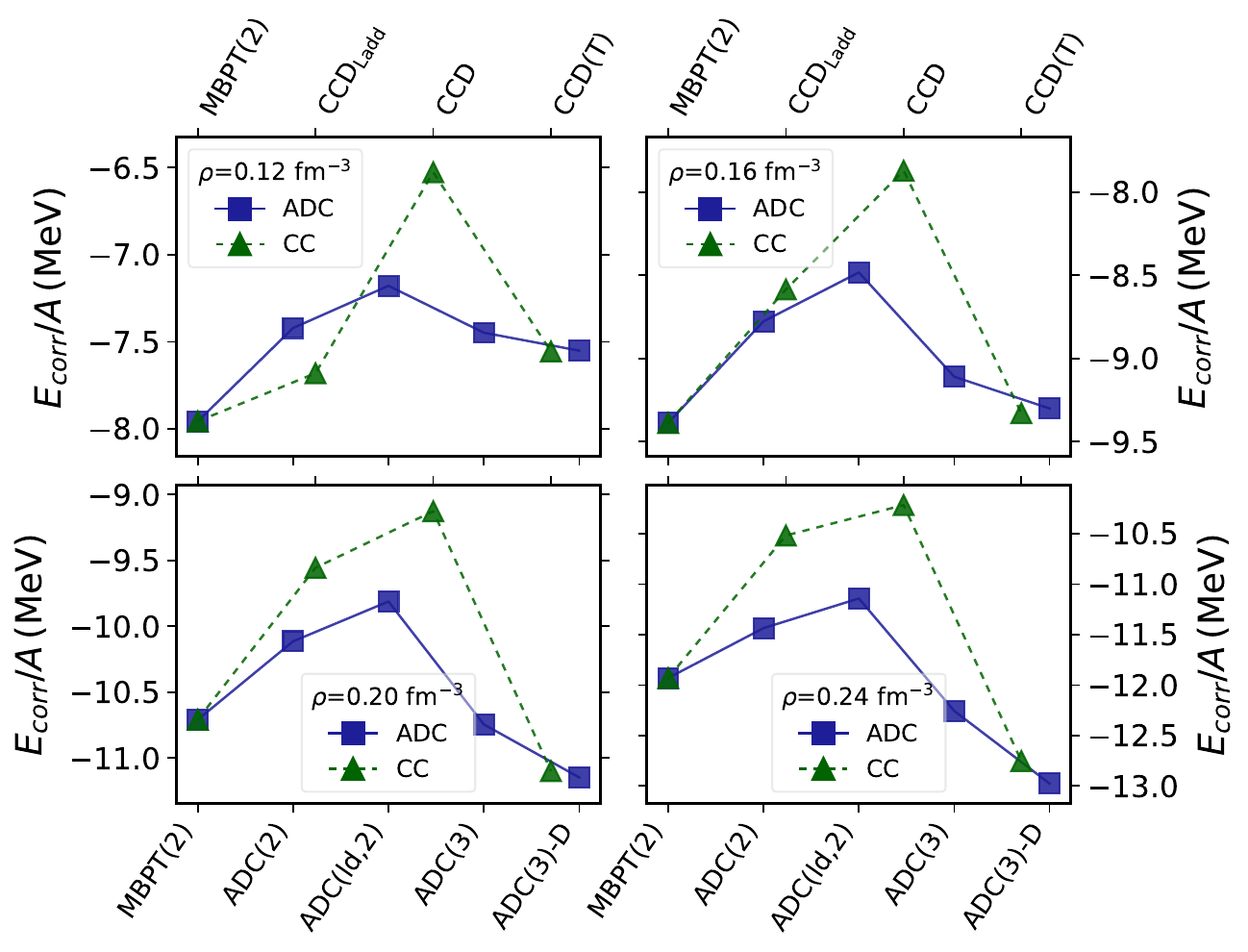}
    \caption{
    Correlation energy per particle in SNM at various densities as obtained by different methods using the $\rm{ NNLO_{sat} }(450)$ interaction.
    Calculations with CC (ADC) schemes of increasing complexity are shown as green triangles (blue squares), and the approximation names are reported in the top (bottom) horizontal axis.
    MBPT(2) is taken as the first point of both the CC and ADC sequences, and ADC(3)-D and CCD(T) results are close to each other to ease their comparison. 
    Lines are a guide to the eye.
    See main text for details.
    }
    \label{fig: levels nnlosat}
\end{figure}

The methods discussed in this work allow in principle to incorporate correlations in a systematic way, which can be represented by the inclusion of appropriate diagrams. 
In this section, we investigate the evolution of the results as a function of the complexity of the approximation scheme.
As an example case, in Fig.~\ref{fig: levels nnlosat} we study the predictions for the correlation energy per particle obtained in SNM with the $\rm{ NNLO_{sat} }(450)$ potential at four densities between 0.12 and 0.24 $\rm{fm}^{-3}$.
In each panel, the sequence of CC (ADC) approximations is shown as green triangles (blue squares).
For both CC and ADC, the first point is taken to be MBPT(2), which provides a baseline reference.
The CC (ADC) variants are labeled in the top (bottom) horizontal axis.
The endpoints for CC and ADC, which are represented by CCD(T) and ADC(3)-D, respectively, are drawn close to each other, to emphasize that they include a similar set of correlations.
Note, though, that there is no precise one-to-one correspondence between the two classes of truncations. 

The following intermediate approximations are introduced. 
"$\rm{ CCD_{Ladd} }$" denotes the CC variant presented e.g. in Refs.~\cite{Baardsen2013,LietzCompNucl}, where the $T_2$ amplitudes are determined as the solution to a simplification of CCD equations, retaining the linear $pp$ and $hh$ terms while neglecting the terms quadratic in $T_2$ and the linear $ph$ rings (see Fig.~\ref{fig: CCD equations}).
The ADC(2) approximation is defined by the standard second-order self-energy diagram (first diagram of Fig.~\ref{fig: ADC(3) diags}).
Finally, "ADC(ld,2)" is defined by an all-order ladder resummation~\cite{MarinoPhdThesis}. More precisely, in this scheme the coupling matrices are treated as in ADC(2), while the interaction matrices feature the $pp$/$hh$ contributions from the third-order ladder diagram (second diagram in Fig.~\ref{fig: ADC(3) diags}).

ADC(2) typically lacks some correlation energy, with the exception of $\rho=0.12\,\rm{fm}^{-3}$, and is higher in energy than ADC(3) by about 0.5 MeV on average.
ADC(ld,2) adds a further repulsive contribution to ADC(2), and thus tends to worsen the agreement with the more refined ADC(3) and CCD(T) schemes. Previous calculations~\cite{Baldo2012,Baardsen2013} pointed out a tendency of ladder GFs to underestimate binding energies in nuclear matter and ascribed this effect to the inclusion of $hh$ scattering terms.
Thus, $ph$ rings are non-negligible according to our results, and we note that, while ADC(3) represents an improvement over ADC(2), \textit{a priori} that may be not the case for approximations that do not follow the strict ADC hierarchy. 
Finally, ADC(3)-D consistenly yields lower energy than ADC(3), with the magnitude of the deviation increasing with the density.

As far as coupled-cluster is concerned, $\rm{ CCD_{Ladd} }$ results are intermediate between MBPT(2) and CCD, but move gradually closer to the CCD ones as the density increases. 
Therefore, the relative contribution of the ladder diagrams to the CC energy grows as a function of the density, and at $\rho=0.24\,\rm{fm}^{-3}$ accounts for most of the correlation energy.
This differs from the homogeneous electron gas (HEG)~\cite{Shepherd2014}, in which ladder (ring) diagrams alone tend to be more repulsive (attractive) than CCD. The physics is also different, though, as rings are the dominant contribution in the HEG, while ladders are in nuclear matter. 
It is also important to note that the excess repulsion of CCD is compensated in CCD(T) by a significant density-dependence of the correction~\eqref{eq: E triples}, which almost doubles from -1 MeV at $\rho=0.08\,\rm{fm}^{-3}$ to -2 MeV at $\rho=0.24\,\rm{fm}^{-3}$. 

\begin{figure}[h!]
    \centering
    \includegraphics[width=\columnwidth]{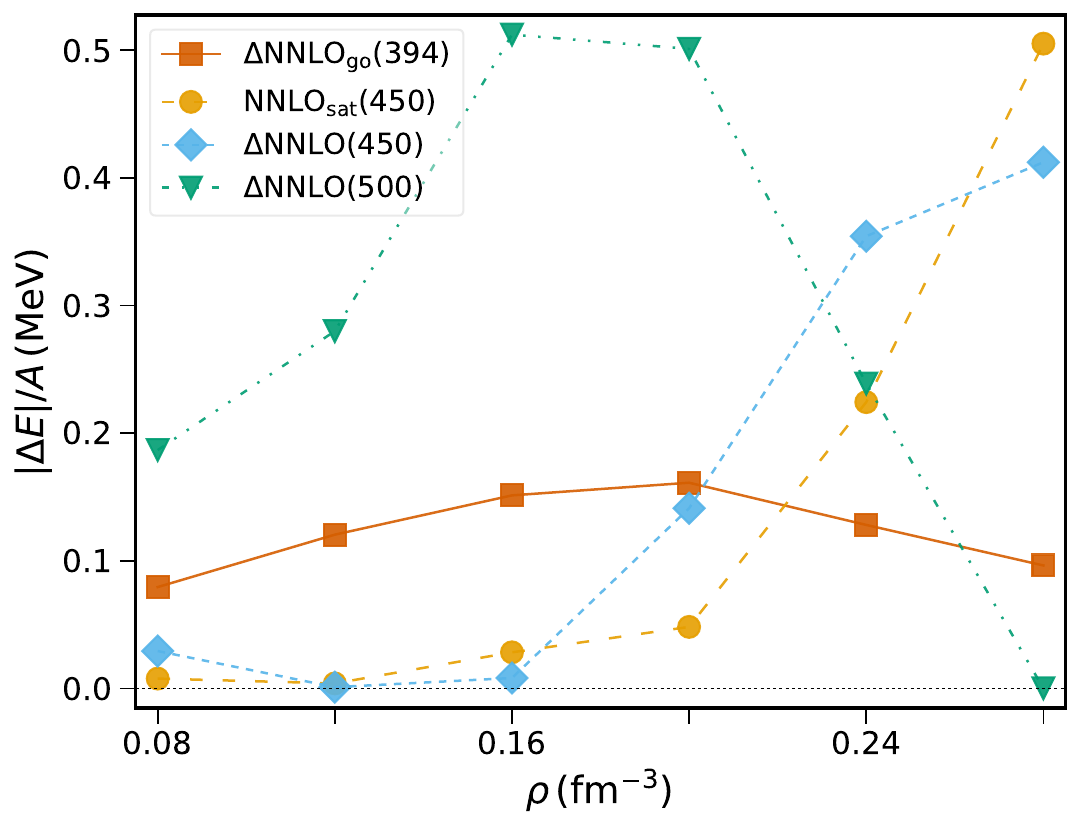}
    \caption{
    Difference (in absolute value) between ACD(3)-D and CCD(T) energies per particle in SNM for different chiral interactions (see legend).
    }
    \label{fig: snm diff}
\end{figure}

%To summarize, g 
General trends have been observed by comparing different CC and ADC truncation schemes. 
CCD(T) and ADC(3) agree well, with CCD(T) always yielding additional binding energy. 
This is consistent with the situation in nuclei, where numerical calculations show that the agreement of Dyson-ADC(3), CC with perturbative triples, as well as in-medium SRG at the two-body level [IM-SRG(2)] with perturbative $3p3h$ correction~\cite{Hergert2020}, all containing similar correlations, is generally good.
Importantly, ADC(3)-D is in impressive quantitative accordance with CCD(T). 
The ability of this scheme to handle strong correlations, already noted in the past~\cite{Barbieri2017,Soma2013}, is now confirmed by our extensive nuclear matter calculations, and the excellent agreement of ADC(3)-D and CCD(T) is summarized in Fig.~\ref{fig: snm diff}, in which the absolute difference between their predictions for the SNM EOS is reported for different interactions.
Discrepancies lie within a few hundred keV, but, with the exception of $\rm{ \Delta NNLO (500)}$, are smaller than $200\,\rm{keV}$ up to $\rho=0.20\,\rm{fm}^{-3}$, well above saturation density.  

The pattern $"\rm{ADC(3)} > \rm{ADC(3)-D} \approx \rm{ CCD(T) }"$ is established for the SNM (and PNM) energies per particles across different potentials and density ranges.
ADC(2,ld), $\rm{ CCD_{Ladd} }$, and especially CCD underestimate correlation energies. 
The relatively simple ADC(2) also provides less binding than ADC(3), but is overall more accurate and consistent in its trend than the previous ones, and, due to its computational lightness, it is suggested as a fast way of obtaining upper bounds on the nuclear matter EOS.

\subsection{Nuclear matter saturation point}
\label{sec: saturation}
\begin{figure}[ht!]
    \centering
    \includegraphics[width=\columnwidth]{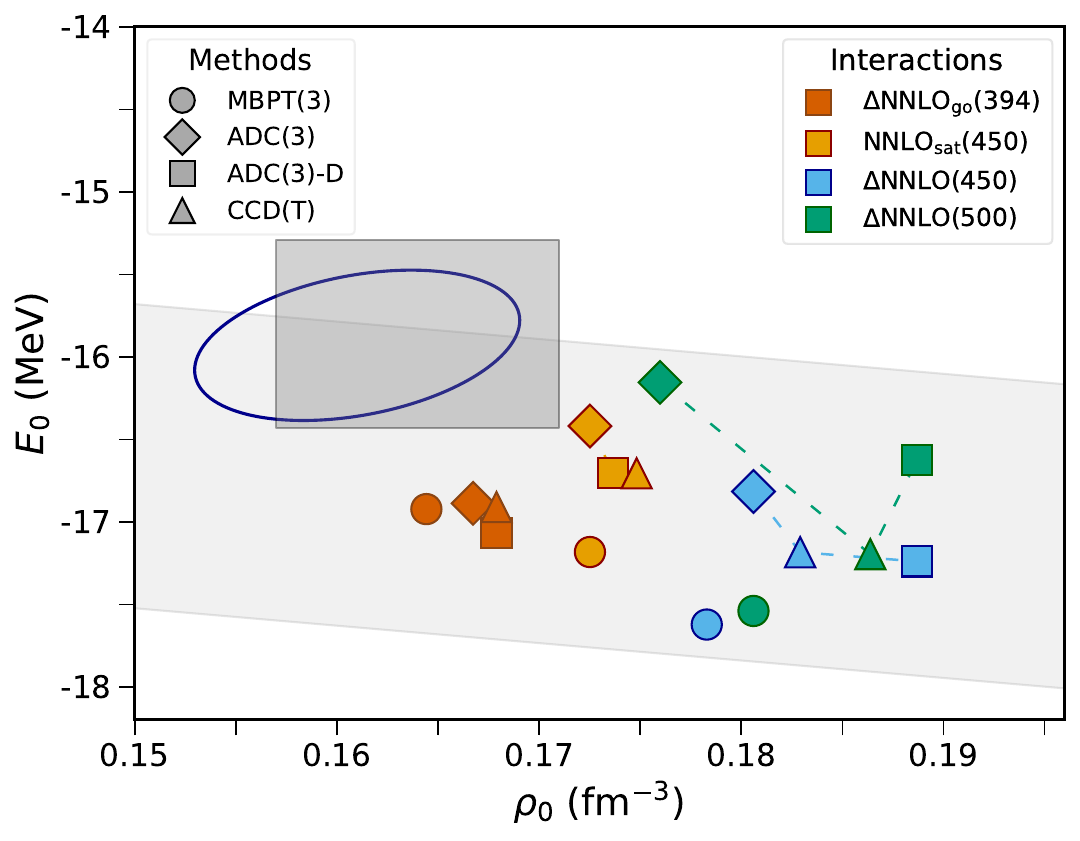}
    \caption{Saturation points predicted in SNM by several interactions (represented with different colors) using different many-body methods (different markers). 
    The grey box represents the empirical saturation region from Ref.~\cite{Drischler2021Review}, and the blue ellipse the 95\% confidence region for the Bayesian posterior distributions for the saturation point estimated from a set of DFT predictions in Ref.~\cite{Drischler2024Bayesian} (left panel of Fig. (5) in~\cite{Drischler2024Bayesian}).
    The saturation points are aligned along a Coester-like line, that is shown as a light-grey anticorrelation band.
    Results for the non-perturbative ADC(3), ADC(3)-D, and CCD(T) calculations are connected by dashed lines.
    }
    \label{fig: saturation snm}
\end{figure}

Finally, in Fig.~\ref{fig: saturation snm} predictions of the SNM saturation point for several chiral interactions and methods are reported.
The saturation energy $E_{0}$ is plotted versus the saturation density $\rho_{0}$, where colors (markers) denote different potentials (techniques).
The position of the saturation point was extracted by interpolating the EOS with a cubic spline and then searching for its minimum. This is an accurate estimate as calculations have been performed over a fine density mesh.
The grey box depicts the estimate for the empirical saturation point taken from Ref.~\cite{Drischler2021Review}, $\rho_{0} = 0.164 \pm 0.007\,\rm{fm}^{-3}$ and $E_{0} = - 15.86 \pm 0.57 \,\rm{MeV}$, based on a selection of density functional theory (DFT)~\cite{colo2020,Schunck2019} predictions.
Also, a Bayesian analysis proposed recently in Ref.~\cite{Drischler2024Bayesian} estimates posterior distributions for the saturation parameters using an extensive set of non-relativistic DFT models.
Thus, we include a blue ellipse (plotted using the code from~\cite{saturationGitHub}), which corresponds to the credibility region associated with the 95\% confidence level shown in ~\cite{Drischler2024Bayesian}, Fig. (5) (left panel) and Eq. (23).
In addition, a Coester-like band encompassing all results is shown as a grey shaded area, with its slope determined by a simple linear fit to the data points.
The outcomes of the non-perturbative calculations - ADC(3), ADC(3)-D, and CCD(T) - are connected by a dashed line for each potential separately to highlight that $\rho_{0}$ and $E_{0}$ are strongly anticorrelated.
MBPT results are anticorrelated with each other, but their trend differs somewhat from that of non-perturbative methods.
In particular, lower saturation energies are predicted, but saturation densities are intermediate between those of ADC and CCD(T).

Overall, the results of Fig.~\ref{fig: saturation snm} allow us to gauge the accuracy that can be presently obtained with \textit{ab initio} techniques for the saturation point of nuclear matter.
As mentioned before, the spreading between CCD(T) and ADC results, particularly between ADC(3) and CCD(T), can be taken as a rough measure of the method error involved in solving the many-nucleon \Sch equation and varies significantly according to the employed interaction. 
Predictions for the low-momentum $\rm{ \Delta NNLO_{go} (394)}$ potential are very well constrained, showing substantial agreement even with MBPT(3).
Tight predictions are also obtained for the two forces with cutoff at 450 $\rm{MeV/c}$. Note that our new calculations show that $\rm{ \Delta NNLO_{go} (394)}$ and $\rm{ NNLO_{sat} }(450)$ saturate rather close to, or even within, the empirical saturation region, as represented by the grey box or the right edge of the blue ellipse. 
The apparent puzzle regarding $\rm{ NNLO_{sat} }(450)$ can be now considered solved in light of these results (see also~\cite{NNLOsatErratum,PbAbInitioAuthorCorrection}). $\rm{ NNLO_{sat} }(450)$ is indeed accurate at the same time for both bulk properties of nuclei and nuclear matter.

All the chiral interactions we have considered saturate at densities $\ge0.16\,\rm{fm}^{-3}$, showing a systematic tendency to overestimate the empirical saturation density.
This may be connected to a tendency to underestimate charge radii in medium-mass nuclei (see e.g.~\cite{Soma2020Chiral,Hergert2020,DeltaGo2020}), though this is not universal, as shown by $\rm{ NNLO_{sat} }$ in calcium~\cite{GarciaRuiz2016} and potassium isotopes~\cite{Koszorus2020}. 
This may be due to the contact nature of the three-nucleon force used in the current truncation of the EFT expansion~\cite{Hebeler3nf}. To achieve a satisfactory description of the EOS at various densities, a higher-order 3N force with a finite range might be necessary. It is worth noting that for DFT interactions (consistent with the empirical saturation region), the density-dependent term simulating the repulsive 3N forces exhibits a more flexible density dependence~\cite{colo2020}.
This means the interactions can be precisely adjusted to match a given saturation point while still providing accurate predictions for finite nuclei observables. However, this is not the case for the chiral interactions used in this work.

\section{Conclusions}
\label{sec: conclusions}

In this work, we have investigated the equations of state of SNM and PNM obtained with a selection of chiral interactions at NNLO, including both NN and 3N forces.
We have employed three diagrammatic many-body methods: CC, ADC-SCGF, and MBPT, and compared the predictions of different truncation schemes of increasing complexity.

Computations in PNM show an essential agreement between the three methods, with differences that amount to a few hundred keV at most.
Thus, we can conclude that, as far as the solution of the \Sch equation is concerned, PNM is rather well controlled, and accurate predictions can be obtained.
On the other hand, the dependence of the PNM EOS on the underlying nuclear force is significant.
Although no systematic study has been attempted here, we have noted that the issue of an unrealistic symmetry energy of $\rm{ NNLO_{sat} }$ has been partially corrected by the delta-full potentials $\rm{ \Delta NNLO }$~\cite{Ekstrom2018Delta} and $\rm{ \Delta NNLO_{go} }$~\cite{DeltaGo2020}. 
These produce a stiffer EOS near saturation density and are generally more consistent with each other in this region.

SNM is a strongly-correlated fermion system and presents a tougher problem from the many-body theory perspective. 
Accordingly, discrepancies between methods are expected to be larger than in PNM.
Nonetheless, our study suggests that, with the partial exception of the difficult case of the hard $\rm{\Delta NNLO(500) }$ interaction, a rather satisfying quantitative agreement can be reached between CCD(T), ADC(3), and ADC(3)-D over a wide range of nucleon densities.
A systematic pattern is recognized, in which CCD(T) provides the lowest energies per particle and ADC(3) underbinds CCD(T) by a small amount.
ADC(3)-D generally compares extremely well with CCD(T), and slight deviations manifest themselves only when correlations are very strong, e.g. with a hard Hamiltonian at $\rho > \rho_0$.
MBPT(3) can be accurate with very soft potentials, such as $\rm{\Delta NNLO_{go}(394) }$; however, it generally tends to be overly attractive at intermediate densities, but more repulsive than CCD(T) at larger densities. Thus, non-perturbative techniques appear superior to MBPT(3) for a quantitative description of SNM.

The robustness of our results is exemplified by considering the SNM saturation point. The outcomes of CC and ADC for both the saturation energy and density are consistent with each other, and in at least one case [$\rm{\Delta NNLO_{go}(394) }$], they almost overlap. 
Moreover, the two quantities exhibit an approximately linear trend which also extends across different interactions.
Finally, our calculations highlight that $\rm{\Delta NNLO_{go}(394) }$ and $\rm{ NNLO_{sat} (450)}$ are reasonably compatible with the empirical constraints on the saturation point.
Thus, these two models manage to simultaneously reproduce SNM properties and bulk observables of finite nuclei, and a puzzle concerning the apparent underbinding~\cite{NNLOsatErratum} of $E_{0}$ by $\rm{ NNLO_{sat}}$ has now been solved. 

To summarize, our work provides accurate estimates of the PNM and SNM EOS predicted by chiral interactions, particularly with soft potentials, as the result of a detailed comparison of diagrammatic \textit{ab initio} approaches.
Our calculations can serve as a benchmark for future studies of nuclear matter, and firmly establish ADC and CC as valuable tools for this purpose.

We believe that these benchmarks between different \textit{ab initio} approaches can provide valuable insights for incorporating systematic many-body correlations with diagrammatic expansion and can pave the way for future studies aimed at pushing many-body method truncations to higher orders.

\section*{Acknowledgements}
We are grateful to  Carlo Barbieri and Gianluca Col\`{o} for constructive feedback on the manuscript.
We also thank Gaute Hagen for fruitful discussions, Y. Dietz for help in benchmarking MBPT(3), and Arnau Rios for further checks on our results.
F.M. acknowledges the CINECA awards AbINEF (HP10B3BG09) and RespGF (HP10BQMECT) under the ISCRA initiative, for the availability of high-performance computing resources and support.
Part of the calculations were also performed using HPC resources at the DiRAC DiAL system at the University of Leicester, United Kingdom (BIS National E-infrastructure Capital Grant No. ST/K000373/1 and STFC Grant No. ST/K0003259/1) and at the supercomputer Mogon at Johannes Gutenberg Universit\"{a}t Mainz.
S.J.N. acknowledges the support from the U.S. Department of Energy (DOE), Office of Science, under SciDAC-5 (NUCLEI collaboration), under grant DE-FG02-97ER41014, and computing resources from the Oak Ridge Leadership Computing Facility located at Oak Ridge National Laboratory, which is supported under contract No. DE-AC05-00OR22725.
This work is supported by the Deutsche Forschungsgemeinschaft (DFG) through Project-ID 279384907 - SFB 1245 and through the Cluster of Excellence “Precision Physics, Fundamental Interactions, and Structure of Matter” (PRISMA+ EXC 2118/1) funded by the DFG within the German Excellence Strategy (Project ID 39083149).

\bibliography{bibliography.bib} 

\begin{thebibliography}{93}%
\makeatletter
\providecommand \@ifxundefined [1]{%
 \@ifx{#1\undefined}
}%
\providecommand \@ifnum [1]{%
 \ifnum #1\expandafter \@firstoftwo
 \else \expandafter \@secondoftwo
 \fi
}%
\providecommand \@ifx [1]{%
 \ifx #1\expandafter \@firstoftwo
 \else \expandafter \@secondoftwo
 \fi
}%
\providecommand \natexlab [1]{#1}%
\providecommand \enquote  [1]{``#1''}%
\providecommand \bibnamefont  [1]{#1}%
\providecommand \bibfnamefont [1]{#1}%
\providecommand \citenamefont [1]{#1}%
\providecommand \href@noop [0]{\@secondoftwo}%
\providecommand \href [0]{\begingroup \@sanitize@url \@href}%
\providecommand \@href[1]{\@@startlink{#1}\@@href}%
\providecommand \@@href[1]{\endgroup#1\@@endlink}%
\providecommand \@sanitize@url [0]{\catcode `\\12\catcode `\$12\catcode `\&12\catcode `\#12\catcode `\^12\catcode `\_12\catcode `\%12\relax}%
\providecommand \@@startlink[1]{}%
\providecommand \@@endlink[0]{}%
\providecommand \url  [0]{\begingroup\@sanitize@url \@url }%
\providecommand \@url [1]{\endgroup\@href {#1}{\urlprefix }}%
\providecommand \urlprefix  [0]{URL }%
\providecommand \Eprint [0]{\href }%
\providecommand \doibase [0]{http://dx.doi.org/}%
\providecommand \selectlanguage [0]{\@gobble}%
\providecommand \bibinfo  [0]{\@secondoftwo}%
\providecommand \bibfield  [0]{\@secondoftwo}%
\providecommand \translation [1]{[#1]}%
\providecommand \BibitemOpen [0]{}%
\providecommand \bibitemStop [0]{}%
\providecommand \bibitemNoStop [0]{.\EOS\space}%
\providecommand \EOS [0]{\spacefactor3000\relax}%
\providecommand \BibitemShut  [1]{\csname bibitem#1\endcsname}%
\let\auto@bib@innerbib\@empty
%</preamble>
\bibitem [{\citenamefont {Burgio}\ \emph {et~al.}(2021)\citenamefont {Burgio}, \citenamefont {Schulze}, \citenamefont {Vidaña},\ and\ \citenamefont {Wei}}]{Burgio2021}%
  \BibitemOpen
  \bibfield  {author} {\bibinfo {author} {\bibfnamefont {G.}~\bibnamefont {Burgio}}, \bibinfo {author} {\bibfnamefont {H.-J.}\ \bibnamefont {Schulze}}, \bibinfo {author} {\bibfnamefont {I.}~\bibnamefont {Vidaña}}, \ and\ \bibinfo {author} {\bibfnamefont {J.-B.}\ \bibnamefont {Wei}},\ }\href {https://www.sciencedirect.com/science/article/pii/S0146641021000338} {\bibfield  {journal} {\bibinfo  {journal} {Progress in Particle and Nuclear Physics}\ }\textbf {\bibinfo {volume} {120}},\ \bibinfo {pages} {103879} (\bibinfo {year} {2021})}\BibitemShut {NoStop}%
\bibitem [{\citenamefont {Haensel}\ \emph {et~al.}(2007)\citenamefont {Haensel}, \citenamefont {Potekhin},\ and\ \citenamefont {Yakovlev}}]{HaenselNeutronStars}%
  \BibitemOpen
  \bibfield  {author} {\bibinfo {author} {\bibfnamefont {P.}~\bibnamefont {Haensel}}, \bibinfo {author} {\bibfnamefont {A.~Y.}\ \bibnamefont {Potekhin}}, \ and\ \bibinfo {author} {\bibfnamefont {D.~G.}\ \bibnamefont {Yakovlev}},\ }\href {\doibase 10.1007/978-0-387-47301-7} {\emph {\bibinfo {title} {{Neutron stars 1: Equation of state and structure}}}},\ Vol.\ \bibinfo {volume} {326}\ (\bibinfo  {publisher} {Springer},\ \bibinfo {address} {New York, USA},\ \bibinfo {year} {2007})\BibitemShut {NoStop}%
\bibitem [{\citenamefont {Sumiyoshi}\ \emph {et~al.}(2023)\citenamefont {Sumiyoshi}, \citenamefont {Kojo},\ and\ \citenamefont {Furusawa}}]{Sumiyoshi2023}%
  \BibitemOpen
  \bibfield  {author} {\bibinfo {author} {\bibfnamefont {K.}~\bibnamefont {Sumiyoshi}}, \bibinfo {author} {\bibfnamefont {T.}~\bibnamefont {Kojo}}, \ and\ \bibinfo {author} {\bibfnamefont {S.}~\bibnamefont {Furusawa}},\ }\enquote {\bibinfo {title} {Equation of state in neutron stars and supernovae},}\ in\ \href {\doibase 10.1007/978-981-19-6345-2_104} {\emph {\bibinfo {booktitle} {Handbook of Nuclear Physics}}},\ \bibinfo {editor} {edited by\ \bibinfo {editor} {\bibfnamefont {I.}~\bibnamefont {Tanihata}}, \bibinfo {editor} {\bibfnamefont {H.}~\bibnamefont {Toki}}, \ and\ \bibinfo {editor} {\bibfnamefont {T.}~\bibnamefont {Kajino}}}\ (\bibinfo  {publisher} {Springer Nature Singapore},\ \bibinfo {address} {Singapore},\ \bibinfo {year} {2023})\ pp.\ \bibinfo {pages} {3127--3177}\BibitemShut {NoStop}%
\bibitem [{\citenamefont {Chatziioannou}\ \emph {et~al.}(2024)\citenamefont {Chatziioannou}, \citenamefont {Cromartie}, \citenamefont {Gandolfi}, \citenamefont {Tews}, \citenamefont {Radice}, \citenamefont {Steiner},\ and\ \citenamefont {Watts}}]{chatziioannou2024neutronstarsdensematter}%
  \BibitemOpen
  \bibfield  {author} {\bibinfo {author} {\bibfnamefont {K.}~\bibnamefont {Chatziioannou}}, \bibinfo {author} {\bibfnamefont {H.~T.}\ \bibnamefont {Cromartie}}, \bibinfo {author} {\bibfnamefont {S.}~\bibnamefont {Gandolfi}}, \bibinfo {author} {\bibfnamefont {I.}~\bibnamefont {Tews}}, \bibinfo {author} {\bibfnamefont {D.}~\bibnamefont {Radice}}, \bibinfo {author} {\bibfnamefont {A.~W.}\ \bibnamefont {Steiner}}, \ and\ \bibinfo {author} {\bibfnamefont {A.~L.}\ \bibnamefont {Watts}},\ }\href {https://arxiv.org/abs/2407.11153} {\enquote {\bibinfo {title} {Neutron stars and the dense matter equation of state: from microscopic theory to macroscopic observations},}\ } (\bibinfo {year} {2024}),\ \Eprint {http://arxiv.org/abs/2407.11153} {arXiv:2407.11153 [nucl-th]} \BibitemShut {NoStop}%
\bibitem [{\citenamefont {Watts}\ \emph {et~al.}(2016)\citenamefont {Watts}, \citenamefont {Andersson}, \citenamefont {Chakrabarty}, \citenamefont {Feroci}, \citenamefont {Hebeler}, \citenamefont {Israel}, \citenamefont {Lamb}, \citenamefont {Miller}, \citenamefont {Morsink}, \citenamefont {\"Ozel}, \citenamefont {Patruno}, \citenamefont {Poutanen}, \citenamefont {Psaltis}, \citenamefont {Schwenk}, \citenamefont {Steiner}, \citenamefont {Stella}, \citenamefont {Tolos},\ and\ \citenamefont {van~der Klis}}]{Watts2016}%
  \BibitemOpen
  \bibfield  {author} {\bibinfo {author} {\bibfnamefont {A.~L.}\ \bibnamefont {Watts}}, \bibinfo {author} {\bibfnamefont {N.}~\bibnamefont {Andersson}}, \bibinfo {author} {\bibfnamefont {D.}~\bibnamefont {Chakrabarty}}, \bibinfo {author} {\bibfnamefont {M.}~\bibnamefont {Feroci}}, \bibinfo {author} {\bibfnamefont {K.}~\bibnamefont {Hebeler}}, \bibinfo {author} {\bibfnamefont {G.}~\bibnamefont {Israel}}, \bibinfo {author} {\bibfnamefont {F.~K.}\ \bibnamefont {Lamb}}, \bibinfo {author} {\bibfnamefont {M.~C.}\ \bibnamefont {Miller}}, \bibinfo {author} {\bibfnamefont {S.}~\bibnamefont {Morsink}}, \bibinfo {author} {\bibfnamefont {F.}~\bibnamefont {\"Ozel}}, \bibinfo {author} {\bibfnamefont {A.}~\bibnamefont {Patruno}}, \bibinfo {author} {\bibfnamefont {J.}~\bibnamefont {Poutanen}}, \bibinfo {author} {\bibfnamefont {D.}~\bibnamefont {Psaltis}}, \bibinfo {author} {\bibfnamefont {A.}~\bibnamefont {Schwenk}}, \bibinfo {author} {\bibfnamefont {A.~W.}\ \bibnamefont {Steiner}}, \bibinfo {author} {\bibfnamefont
  {L.}~\bibnamefont {Stella}}, \bibinfo {author} {\bibfnamefont {L.}~\bibnamefont {Tolos}}, \ and\ \bibinfo {author} {\bibfnamefont {M.}~\bibnamefont {van~der Klis}},\ }\href {\doibase 10.1103/RevModPhys.88.021001} {\bibfield  {journal} {\bibinfo  {journal} {Rev. Mod. Phys.}\ }\textbf {\bibinfo {volume} {88}},\ \bibinfo {pages} {021001} (\bibinfo {year} {2016})}\BibitemShut {NoStop}%
\bibitem [{\citenamefont {Baiotti}(2019)}]{Baiotti2019}%
  \BibitemOpen
  \bibfield  {author} {\bibinfo {author} {\bibfnamefont {L.}~\bibnamefont {Baiotti}},\ }\href {\doibase https://doi.org/10.1016/j.ppnp.2019.103714} {\bibfield  {journal} {\bibinfo  {journal} {Progress in Particle and Nuclear Physics}\ }\textbf {\bibinfo {volume} {109}},\ \bibinfo {pages} {103714} (\bibinfo {year} {2019})}\BibitemShut {NoStop}%
\bibitem [{\citenamefont {Annala}\ \emph {et~al.}(2022)\citenamefont {Annala}, \citenamefont {Gorda}, \citenamefont {Katerini}, \citenamefont {Kurkela}, \citenamefont {N\"attil\"a}, \citenamefont {Paschalidis},\ and\ \citenamefont {Vuorinen}}]{Gorda2022}%
  \BibitemOpen
  \bibfield  {author} {\bibinfo {author} {\bibfnamefont {E.}~\bibnamefont {Annala}}, \bibinfo {author} {\bibfnamefont {T.}~\bibnamefont {Gorda}}, \bibinfo {author} {\bibfnamefont {E.}~\bibnamefont {Katerini}}, \bibinfo {author} {\bibfnamefont {A.}~\bibnamefont {Kurkela}}, \bibinfo {author} {\bibfnamefont {J.}~\bibnamefont {N\"attil\"a}}, \bibinfo {author} {\bibfnamefont {V.}~\bibnamefont {Paschalidis}}, \ and\ \bibinfo {author} {\bibfnamefont {A.}~\bibnamefont {Vuorinen}},\ }\href {\doibase 10.1103/PhysRevX.12.011058} {\bibfield  {journal} {\bibinfo  {journal} {Phys. Rev. X}\ }\textbf {\bibinfo {volume} {12}},\ \bibinfo {pages} {011058} (\bibinfo {year} {2022})}\BibitemShut {NoStop}%
\bibitem [{\citenamefont {Drischler}\ \emph {et~al.}(2021)\citenamefont {Drischler}, \citenamefont {Holt},\ and\ \citenamefont {Wellenhofer}}]{Drischler2021Review}%
  \BibitemOpen
  \bibfield  {author} {\bibinfo {author} {\bibfnamefont {C.}~\bibnamefont {Drischler}}, \bibinfo {author} {\bibfnamefont {J.}~\bibnamefont {Holt}}, \ and\ \bibinfo {author} {\bibfnamefont {C.}~\bibnamefont {Wellenhofer}},\ }\href {\doibase 10.1146/annurev-nucl-102419-041903} {\bibfield  {journal} {\bibinfo  {journal} {Annual Review of Nuclear and Particle Science}\ }\textbf {\bibinfo {volume} {71}},\ \bibinfo {pages} {403} (\bibinfo {year} {2021})}\BibitemShut {NoStop}%
\bibitem [{\citenamefont {Huth}\ \emph {et~al.}(2022)\citenamefont {Huth}, \citenamefont {Pang}, \citenamefont {Tews}, \citenamefont {Dietrich}, \citenamefont {Le~F{\`e}vre}, \citenamefont {Schwenk}, \citenamefont {Trautmann}, \citenamefont {Agarwal}, \citenamefont {Bulla}, \citenamefont {Coughlin} \emph {et~al.}}]{Huth2022}%
  \BibitemOpen
  \bibfield  {author} {\bibinfo {author} {\bibfnamefont {S.}~\bibnamefont {Huth}}, \bibinfo {author} {\bibfnamefont {P.~T.}\ \bibnamefont {Pang}}, \bibinfo {author} {\bibfnamefont {I.}~\bibnamefont {Tews}}, \bibinfo {author} {\bibfnamefont {T.}~\bibnamefont {Dietrich}}, \bibinfo {author} {\bibfnamefont {A.}~\bibnamefont {Le~F{\`e}vre}}, \bibinfo {author} {\bibfnamefont {A.}~\bibnamefont {Schwenk}}, \bibinfo {author} {\bibfnamefont {W.}~\bibnamefont {Trautmann}}, \bibinfo {author} {\bibfnamefont {K.}~\bibnamefont {Agarwal}}, \bibinfo {author} {\bibfnamefont {M.}~\bibnamefont {Bulla}}, \bibinfo {author} {\bibfnamefont {M.~W.}\ \bibnamefont {Coughlin}},  \emph {et~al.},\ }\href@noop {} {\bibfield  {journal} {\bibinfo  {journal} {Nature}\ }\textbf {\bibinfo {volume} {606}},\ \bibinfo {pages} {276} (\bibinfo {year} {2022})}\BibitemShut {NoStop}%
\bibitem [{\citenamefont {Simonis}\ \emph {et~al.}(2017)\citenamefont {Simonis}, \citenamefont {Stroberg}, \citenamefont {Hebeler}, \citenamefont {Holt},\ and\ \citenamefont {Schwenk}}]{Simonis2017}%
  \BibitemOpen
  \bibfield  {author} {\bibinfo {author} {\bibfnamefont {J.}~\bibnamefont {Simonis}}, \bibinfo {author} {\bibfnamefont {S.~R.}\ \bibnamefont {Stroberg}}, \bibinfo {author} {\bibfnamefont {K.}~\bibnamefont {Hebeler}}, \bibinfo {author} {\bibfnamefont {J.~D.}\ \bibnamefont {Holt}}, \ and\ \bibinfo {author} {\bibfnamefont {A.}~\bibnamefont {Schwenk}},\ }\href {\doibase 10.1103/PhysRevC.96.014303} {\bibfield  {journal} {\bibinfo  {journal} {Phys. Rev. C}\ }\textbf {\bibinfo {volume} {96}},\ \bibinfo {pages} {014303} (\bibinfo {year} {2017})}\BibitemShut {NoStop}%
\bibitem [{\citenamefont {Machleidt}\ and\ \citenamefont {Sammarruca}(2024)}]{MACHLEIDT2024104117}%
  \BibitemOpen
  \bibfield  {author} {\bibinfo {author} {\bibfnamefont {R.}~\bibnamefont {Machleidt}}\ and\ \bibinfo {author} {\bibfnamefont {F.}~\bibnamefont {Sammarruca}},\ }\href {\doibase https://doi.org/10.1016/j.ppnp.2024.104117} {\bibfield  {journal} {\bibinfo  {journal} {Progress in Particle and Nuclear Physics}\ }\textbf {\bibinfo {volume} {137}},\ \bibinfo {pages} {104117} (\bibinfo {year} {2024})}\BibitemShut {NoStop}%
\bibitem [{\citenamefont {Sammarruca}\ and\ \citenamefont {Millerson}(2020)}]{Sammarruca2020}%
  \BibitemOpen
  \bibfield  {author} {\bibinfo {author} {\bibfnamefont {F.}~\bibnamefont {Sammarruca}}\ and\ \bibinfo {author} {\bibfnamefont {R.}~\bibnamefont {Millerson}},\ }\href {\doibase 10.1103/PhysRevC.102.034313} {\bibfield  {journal} {\bibinfo  {journal} {Phys. Rev. C}\ }\textbf {\bibinfo {volume} {102}},\ \bibinfo {pages} {034313} (\bibinfo {year} {2020})}\BibitemShut {NoStop}%
\bibitem [{\citenamefont {Ekstr\"om}\ \emph {et~al.}(2015)\citenamefont {Ekstr\"om}, \citenamefont {Jansen}, \citenamefont {Wendt}, \citenamefont {Hagen}, \citenamefont {Papenbrock}, \citenamefont {Carlsson}, \citenamefont {Forss\'en}, \citenamefont {Hjorth-Jensen}, \citenamefont {Navr\'atil},\ and\ \citenamefont {Nazarewicz}}]{NNLOsat}%
  \BibitemOpen
  \bibfield  {author} {\bibinfo {author} {\bibfnamefont {A.}~\bibnamefont {Ekstr\"om}}, \bibinfo {author} {\bibfnamefont {G.~R.}\ \bibnamefont {Jansen}}, \bibinfo {author} {\bibfnamefont {K.~A.}\ \bibnamefont {Wendt}}, \bibinfo {author} {\bibfnamefont {G.}~\bibnamefont {Hagen}}, \bibinfo {author} {\bibfnamefont {T.}~\bibnamefont {Papenbrock}}, \bibinfo {author} {\bibfnamefont {B.~D.}\ \bibnamefont {Carlsson}}, \bibinfo {author} {\bibfnamefont {C.}~\bibnamefont {Forss\'en}}, \bibinfo {author} {\bibfnamefont {M.}~\bibnamefont {Hjorth-Jensen}}, \bibinfo {author} {\bibfnamefont {P.}~\bibnamefont {Navr\'atil}}, \ and\ \bibinfo {author} {\bibfnamefont {W.}~\bibnamefont {Nazarewicz}},\ }\href {\doibase 10.1103/PhysRevC.91.051301} {\bibfield  {journal} {\bibinfo  {journal} {Phys. Rev. C}\ }\textbf {\bibinfo {volume} {91}},\ \bibinfo {pages} {051301} (\bibinfo {year} {2015})}\BibitemShut {NoStop}%
\bibitem [{\citenamefont {Jiang}\ \emph {et~al.}(2020)\citenamefont {Jiang}, \citenamefont {Ekstr\"om}, \citenamefont {Forss\'en}, \citenamefont {Hagen}, \citenamefont {Jansen},\ and\ \citenamefont {Papenbrock}}]{DeltaGo2020}%
  \BibitemOpen
  \bibfield  {author} {\bibinfo {author} {\bibfnamefont {W.~G.}\ \bibnamefont {Jiang}}, \bibinfo {author} {\bibfnamefont {A.}~\bibnamefont {Ekstr\"om}}, \bibinfo {author} {\bibfnamefont {C.}~\bibnamefont {Forss\'en}}, \bibinfo {author} {\bibfnamefont {G.}~\bibnamefont {Hagen}}, \bibinfo {author} {\bibfnamefont {G.~R.}\ \bibnamefont {Jansen}}, \ and\ \bibinfo {author} {\bibfnamefont {T.}~\bibnamefont {Papenbrock}},\ }\href {\doibase 10.1103/PhysRevC.102.054301} {\bibfield  {journal} {\bibinfo  {journal} {Phys. Rev. C}\ }\textbf {\bibinfo {volume} {102}},\ \bibinfo {pages} {054301} (\bibinfo {year} {2020})}\BibitemShut {NoStop}%
\bibitem [{\citenamefont {Drischler}\ \emph {et~al.}(2019)\citenamefont {Drischler}, \citenamefont {Hebeler},\ and\ \citenamefont {Schwenk}}]{Drischler2019}%
  \BibitemOpen
  \bibfield  {author} {\bibinfo {author} {\bibfnamefont {C.}~\bibnamefont {Drischler}}, \bibinfo {author} {\bibfnamefont {K.}~\bibnamefont {Hebeler}}, \ and\ \bibinfo {author} {\bibfnamefont {A.}~\bibnamefont {Schwenk}},\ }\href {\doibase 10.1103/PhysRevLett.122.042501} {\bibfield  {journal} {\bibinfo  {journal} {Phys. Rev. Lett.}\ }\textbf {\bibinfo {volume} {122}},\ \bibinfo {pages} {042501} (\bibinfo {year} {2019})}\BibitemShut {NoStop}%
\bibitem [{\citenamefont {Jiang}\ \emph {et~al.}(2024)\citenamefont {Jiang}, \citenamefont {Forss\'en}, \citenamefont {Dj\"arv},\ and\ \citenamefont {Hagen}}]{Jiang2024}%
  \BibitemOpen
  \bibfield  {author} {\bibinfo {author} {\bibfnamefont {W.~G.}\ \bibnamefont {Jiang}}, \bibinfo {author} {\bibfnamefont {C.}~\bibnamefont {Forss\'en}}, \bibinfo {author} {\bibfnamefont {T.}~\bibnamefont {Dj\"arv}}, \ and\ \bibinfo {author} {\bibfnamefont {G.}~\bibnamefont {Hagen}},\ }\href {\doibase 10.1103/PhysRevC.109.L061302} {\bibfield  {journal} {\bibinfo  {journal} {Phys. Rev. C}\ }\textbf {\bibinfo {volume} {109}},\ \bibinfo {pages} {L061302} (\bibinfo {year} {2024})}\BibitemShut {NoStop}%
\bibitem [{\citenamefont {Hu}\ \emph {et~al.}(2022)\citenamefont {Hu}, \citenamefont {Jiang}, \citenamefont {Miyagi}, \citenamefont {Sun}, \citenamefont {Ekström}, \citenamefont {Forssén}, \citenamefont {Hagen}, \citenamefont {Holt}, \citenamefont {Papenbrock}, \citenamefont {Stroberg},\ and\ \citenamefont {Vernon}}]{PbAbInitio}%
  \BibitemOpen
  \bibfield  {author} {\bibinfo {author} {\bibfnamefont {B.}~\bibnamefont {Hu}}, \bibinfo {author} {\bibfnamefont {W.}~\bibnamefont {Jiang}}, \bibinfo {author} {\bibfnamefont {T.}~\bibnamefont {Miyagi}}, \bibinfo {author} {\bibfnamefont {Z.}~\bibnamefont {Sun}}, \bibinfo {author} {\bibfnamefont {A.}~\bibnamefont {Ekström}}, \bibinfo {author} {\bibfnamefont {C.}~\bibnamefont {Forssén}}, \bibinfo {author} {\bibfnamefont {G.}~\bibnamefont {Hagen}}, \bibinfo {author} {\bibfnamefont {J.~D.}\ \bibnamefont {Holt}}, \bibinfo {author} {\bibfnamefont {T.}~\bibnamefont {Papenbrock}}, \bibinfo {author} {\bibfnamefont {S.~R.}\ \bibnamefont {Stroberg}}, \ and\ \bibinfo {author} {\bibfnamefont {I.}~\bibnamefont {Vernon}},\ }\href {https://www.nature.com/articles/s41567-022-01715-8} {\bibfield  {journal} {\bibinfo  {journal} {Nature Physics}\ }\textbf {\bibinfo {volume} {18}},\ \bibinfo {pages} {1196} (\bibinfo {year} {2022})}\BibitemShut {NoStop}%
\bibitem [{\citenamefont {Gandolfi}\ \emph {et~al.}(2015)\citenamefont {Gandolfi}, \citenamefont {Gezerlis},\ and\ \citenamefont {Carlson}}]{Gandolfi2015}%
  \BibitemOpen
  \bibfield  {author} {\bibinfo {author} {\bibfnamefont {S.}~\bibnamefont {Gandolfi}}, \bibinfo {author} {\bibfnamefont {A.}~\bibnamefont {Gezerlis}}, \ and\ \bibinfo {author} {\bibfnamefont {J.}~\bibnamefont {Carlson}},\ }\href {\doibase 10.1146/annurev-nucl-102014-021957} {\bibfield  {journal} {\bibinfo  {journal} {Annual Review of Nuclear and Particle Science}\ }\textbf {\bibinfo {volume} {65}},\ \bibinfo {pages} {303} (\bibinfo {year} {2015})}\BibitemShut {NoStop}%
\bibitem [{\citenamefont {Piarulli}\ \emph {et~al.}(2020)\citenamefont {Piarulli}, \citenamefont {Bombaci}, \citenamefont {Logoteta}, \citenamefont {Lovato},\ and\ \citenamefont {Wiringa}}]{Piarulli2020}%
  \BibitemOpen
  \bibfield  {author} {\bibinfo {author} {\bibfnamefont {M.}~\bibnamefont {Piarulli}}, \bibinfo {author} {\bibfnamefont {I.}~\bibnamefont {Bombaci}}, \bibinfo {author} {\bibfnamefont {D.}~\bibnamefont {Logoteta}}, \bibinfo {author} {\bibfnamefont {A.}~\bibnamefont {Lovato}}, \ and\ \bibinfo {author} {\bibfnamefont {R.~B.}\ \bibnamefont {Wiringa}},\ }\href {\doibase 10.1103/PhysRevC.101.045801} {\bibfield  {journal} {\bibinfo  {journal} {Phys. Rev. C}\ }\textbf {\bibinfo {volume} {101}},\ \bibinfo {pages} {045801} (\bibinfo {year} {2020})}\BibitemShut {NoStop}%
\bibitem [{\citenamefont {Lovato}\ \emph {et~al.}(2022)\citenamefont {Lovato}, \citenamefont {Bombaci}, \citenamefont {Logoteta}, \citenamefont {Piarulli},\ and\ \citenamefont {Wiringa}}]{Lovato2022PNM}%
  \BibitemOpen
  \bibfield  {author} {\bibinfo {author} {\bibfnamefont {A.}~\bibnamefont {Lovato}}, \bibinfo {author} {\bibfnamefont {I.}~\bibnamefont {Bombaci}}, \bibinfo {author} {\bibfnamefont {D.}~\bibnamefont {Logoteta}}, \bibinfo {author} {\bibfnamefont {M.}~\bibnamefont {Piarulli}}, \ and\ \bibinfo {author} {\bibfnamefont {R.~B.}\ \bibnamefont {Wiringa}},\ }\href {\doibase 10.1103/PhysRevC.105.055808} {\bibfield  {journal} {\bibinfo  {journal} {Phys. Rev. C}\ }\textbf {\bibinfo {volume} {105}},\ \bibinfo {pages} {055808} (\bibinfo {year} {2022})}\BibitemShut {NoStop}%
\bibitem [{\citenamefont {Drischler}\ \emph {et~al.}(2020)\citenamefont {Drischler}, \citenamefont {Furnstahl}, \citenamefont {Melendez},\ and\ \citenamefont {Phillips}}]{Drischler2020}%
  \BibitemOpen
  \bibfield  {author} {\bibinfo {author} {\bibfnamefont {C.}~\bibnamefont {Drischler}}, \bibinfo {author} {\bibfnamefont {R.~J.}\ \bibnamefont {Furnstahl}}, \bibinfo {author} {\bibfnamefont {J.~A.}\ \bibnamefont {Melendez}}, \ and\ \bibinfo {author} {\bibfnamefont {D.~R.}\ \bibnamefont {Phillips}},\ }\href {\doibase 10.1103/PhysRevLett.125.202702} {\bibfield  {journal} {\bibinfo  {journal} {Phys. Rev. Lett.}\ }\textbf {\bibinfo {volume} {125}},\ \bibinfo {pages} {202702} (\bibinfo {year} {2020})}\BibitemShut {NoStop}%
\bibitem [{\citenamefont {Hergert}(2020)}]{Hergert2020}%
  \BibitemOpen
  \bibfield  {author} {\bibinfo {author} {\bibfnamefont {H.}~\bibnamefont {Hergert}},\ }\href {\doibase 10.3389/fphy.2020.00379} {\bibfield  {journal} {\bibinfo  {journal} {Frontiers in Physics}\ }\textbf {\bibinfo {volume} {8}},\ \bibinfo {pages} {379} (\bibinfo {year} {2020})}\BibitemShut {NoStop}%
\bibitem [{\citenamefont {Hjorth-Jensen}\ \emph {et~al.}(2017)\citenamefont {Hjorth-Jensen}, \citenamefont {Lombardo},\ and\ \citenamefont {van Kolck}}]{computational_nuclear}%
  \BibitemOpen
  \bibfield  {author} {\bibinfo {author} {\bibfnamefont {M.}~\bibnamefont {Hjorth-Jensen}}, \bibinfo {author} {\bibfnamefont {M.}~\bibnamefont {Lombardo}}, \ and\ \bibinfo {author} {\bibfnamefont {U.}~\bibnamefont {van Kolck}},\ }\href {\doibase 10.1007/978-3-319-53336-0} {\emph {\bibinfo {title} {An Advanced Course in Computational Nuclear Physics: Bridging the Scales from Quarks to Neutron Stars}}}\ (\bibinfo {year} {2017})\BibitemShut {NoStop}%
\bibitem [{\citenamefont {Piarulli}\ and\ \citenamefont {Tews}(2020)}]{Piarulli2020Delta}%
  \BibitemOpen
  \bibfield  {author} {\bibinfo {author} {\bibfnamefont {M.}~\bibnamefont {Piarulli}}\ and\ \bibinfo {author} {\bibfnamefont {I.}~\bibnamefont {Tews}},\ }\href {https://www.frontiersin.org/articles/10.3389/fphy.2019.00245} {\bibfield  {journal} {\bibinfo  {journal} {Frontiers in Physics}\ }\textbf {\bibinfo {volume} {7}} (\bibinfo {year} {2020})}\BibitemShut {NoStop}%
\bibitem [{\citenamefont {Machleidt}\ and\ \citenamefont {Sammarruca}(2016)}]{Machleidt2016}%
  \BibitemOpen
  \bibfield  {author} {\bibinfo {author} {\bibfnamefont {R.}~\bibnamefont {Machleidt}}\ and\ \bibinfo {author} {\bibfnamefont {F.}~\bibnamefont {Sammarruca}},\ }\href {\doibase 10.1088/0031-8949/91/8/083007} {\bibfield  {journal} {\bibinfo  {journal} {Physica Scripta}\ }\textbf {\bibinfo {volume} {91}},\ \bibinfo {pages} {083007} (\bibinfo {year} {2016})}\BibitemShut {NoStop}%
\bibitem [{\citenamefont {Epelbaum}(2024)}]{Epelbaum2024}%
  \BibitemOpen
  \bibfield  {author} {\bibinfo {author} {\bibfnamefont {E.}~\bibnamefont {Epelbaum}},\ }\href {\doibase 10.1007/s00601-024-01918-0} {\bibfield  {journal} {\bibinfo  {journal} {Few-Body Systems}\ }\textbf {\bibinfo {volume} {65}} (\bibinfo {year} {2024}),\ 10.1007/s00601-024-01918-0}\BibitemShut {NoStop}%
\bibitem [{\citenamefont {Rios}(2020)}]{Rios2020}%
  \BibitemOpen
  \bibfield  {author} {\bibinfo {author} {\bibfnamefont {A.}~\bibnamefont {Rios}},\ }\href {https://www.frontiersin.org/articles/10.3389/fphy.2020.00387} {\bibfield  {journal} {\bibinfo  {journal} {Frontiers in Physics}\ }\textbf {\bibinfo {volume} {8}} (\bibinfo {year} {2020})}\BibitemShut {NoStop}%
\bibitem [{\citenamefont {Carbone}(2020)}]{Carbone2020}%
  \BibitemOpen
  \bibfield  {author} {\bibinfo {author} {\bibfnamefont {A.}~\bibnamefont {Carbone}},\ }\href {\doibase 10.1103/PhysRevResearch.2.023227} {\bibfield  {journal} {\bibinfo  {journal} {Phys. Rev. Research}\ }\textbf {\bibinfo {volume} {2}},\ \bibinfo {pages} {023227} (\bibinfo {year} {2020})}\BibitemShut {NoStop}%
\bibitem [{\citenamefont {Carbone}\ \emph {et~al.}(2014)\citenamefont {Carbone}, \citenamefont {Rios},\ and\ \citenamefont {Polls}}]{Carbone2014}%
  \BibitemOpen
  \bibfield  {author} {\bibinfo {author} {\bibfnamefont {A.}~\bibnamefont {Carbone}}, \bibinfo {author} {\bibfnamefont {A.}~\bibnamefont {Rios}}, \ and\ \bibinfo {author} {\bibfnamefont {A.}~\bibnamefont {Polls}},\ }\href {\doibase 10.1103/PhysRevC.90.054322} {\bibfield  {journal} {\bibinfo  {journal} {Phys. Rev. C}\ }\textbf {\bibinfo {volume} {90}},\ \bibinfo {pages} {054322} (\bibinfo {year} {2014})}\BibitemShut {NoStop}%
\bibitem [{\citenamefont {Carbone}\ \emph {et~al.}(2013)\citenamefont {Carbone}, \citenamefont {Polls},\ and\ \citenamefont {Rios}}]{Carbone2013Sym}%
  \BibitemOpen
  \bibfield  {author} {\bibinfo {author} {\bibfnamefont {A.}~\bibnamefont {Carbone}}, \bibinfo {author} {\bibfnamefont {A.}~\bibnamefont {Polls}}, \ and\ \bibinfo {author} {\bibfnamefont {A.}~\bibnamefont {Rios}},\ }\href {\doibase 10.1103/PhysRevC.88.044302} {\bibfield  {journal} {\bibinfo  {journal} {Phys. Rev. C}\ }\textbf {\bibinfo {volume} {88}},\ \bibinfo {pages} {044302} (\bibinfo {year} {2013})}\BibitemShut {NoStop}%
\bibitem [{\citenamefont {Keller}\ \emph {et~al.}(2023)\citenamefont {Keller}, \citenamefont {Hebeler},\ and\ \citenamefont {Schwenk}}]{Keller2023}%
  \BibitemOpen
  \bibfield  {author} {\bibinfo {author} {\bibfnamefont {J.}~\bibnamefont {Keller}}, \bibinfo {author} {\bibfnamefont {K.}~\bibnamefont {Hebeler}}, \ and\ \bibinfo {author} {\bibfnamefont {A.}~\bibnamefont {Schwenk}},\ }\href {\doibase 10.1103/PhysRevLett.130.072701} {\bibfield  {journal} {\bibinfo  {journal} {Phys. Rev. Lett.}\ }\textbf {\bibinfo {volume} {130}},\ \bibinfo {pages} {072701} (\bibinfo {year} {2023})}\BibitemShut {NoStop}%
\bibitem [{\citenamefont {Tichai}\ \emph {et~al.}(2020)\citenamefont {Tichai}, \citenamefont {Roth},\ and\ \citenamefont {Duguet}}]{Tichai2020Mbpt}%
  \BibitemOpen
  \bibfield  {author} {\bibinfo {author} {\bibfnamefont {A.}~\bibnamefont {Tichai}}, \bibinfo {author} {\bibfnamefont {R.}~\bibnamefont {Roth}}, \ and\ \bibinfo {author} {\bibfnamefont {T.}~\bibnamefont {Duguet}},\ }\href {https://www.frontiersin.org/articles/10.3389/fphy.2020.00164} {\bibfield  {journal} {\bibinfo  {journal} {Frontiers in Physics}\ }\textbf {\bibinfo {volume} {8}} (\bibinfo {year} {2020})}\BibitemShut {NoStop}%
\bibitem [{\citenamefont {Lynn}\ \emph {et~al.}(2019)\citenamefont {Lynn}, \citenamefont {Tews}, \citenamefont {Gandolfi},\ and\ \citenamefont {Lovato}}]{Lynn2019}%
  \BibitemOpen
  \bibfield  {author} {\bibinfo {author} {\bibfnamefont {J.}~\bibnamefont {Lynn}}, \bibinfo {author} {\bibfnamefont {I.}~\bibnamefont {Tews}}, \bibinfo {author} {\bibfnamefont {S.}~\bibnamefont {Gandolfi}}, \ and\ \bibinfo {author} {\bibfnamefont {A.}~\bibnamefont {Lovato}},\ }\href {\doibase 10.1146/annurev-nucl-101918-023600} {\bibfield  {journal} {\bibinfo  {journal} {Annual Review of Nuclear and Particle Science}\ }\textbf {\bibinfo {volume} {69}},\ \bibinfo {pages} {279} (\bibinfo {year} {2019})}\BibitemShut {NoStop}%
\bibitem [{\citenamefont {Roggero}\ \emph {et~al.}(2014)\citenamefont {Roggero}, \citenamefont {Mukherjee},\ and\ \citenamefont {Pederiva}}]{Roggero2014CIMC}%
  \BibitemOpen
  \bibfield  {author} {\bibinfo {author} {\bibfnamefont {A.}~\bibnamefont {Roggero}}, \bibinfo {author} {\bibfnamefont {A.}~\bibnamefont {Mukherjee}}, \ and\ \bibinfo {author} {\bibfnamefont {F.}~\bibnamefont {Pederiva}},\ }\href {\doibase 10.1103/PhysRevLett.112.221103} {\bibfield  {journal} {\bibinfo  {journal} {Phys. Rev. Lett.}\ }\textbf {\bibinfo {volume} {112}},\ \bibinfo {pages} {221103} (\bibinfo {year} {2014})}\BibitemShut {NoStop}%
\bibitem [{\citenamefont {Arthuis}\ \emph {et~al.}(2023)\citenamefont {Arthuis}, \citenamefont {Barbieri}, \citenamefont {Pederiva},\ and\ \citenamefont {Roggero}}]{Arthuis2023}%
  \BibitemOpen
  \bibfield  {author} {\bibinfo {author} {\bibfnamefont {P.}~\bibnamefont {Arthuis}}, \bibinfo {author} {\bibfnamefont {C.}~\bibnamefont {Barbieri}}, \bibinfo {author} {\bibfnamefont {F.}~\bibnamefont {Pederiva}}, \ and\ \bibinfo {author} {\bibfnamefont {A.}~\bibnamefont {Roggero}},\ }\href {https://link.aps.org/doi/10.1103/PhysRevC.107.044303} {\bibfield  {journal} {\bibinfo  {journal} {Phys. Rev. C}\ }\textbf {\bibinfo {volume} {107}},\ \bibinfo {pages} {044303} (\bibinfo {year} {2023})}\BibitemShut {NoStop}%
\bibitem [{\citenamefont {Lonardoni}\ and\ \citenamefont {Tews}(2020)}]{Lonardoni2020LocalChiral}%
  \BibitemOpen
  \bibfield  {author} {\bibinfo {author} {\bibfnamefont {D.}~\bibnamefont {Lonardoni}}\ and\ \bibinfo {author} {\bibfnamefont {I.}~\bibnamefont {Tews}},\ }\href {https://www.osti.gov/biblio/1605124} {\bibfield  {journal} {\bibinfo  {journal} {PoS - Proceedings of Science}\ }\textbf {\bibinfo {volume} {317}} (\bibinfo {year} {2020})}\BibitemShut {NoStop}%
\bibitem [{\citenamefont {Gandolfi}\ \emph {et~al.}(2014)\citenamefont {Gandolfi}, \citenamefont {Lovato}, \citenamefont {Carlson},\ and\ \citenamefont {Schmidt}}]{Gandolfi2014snm}%
  \BibitemOpen
  \bibfield  {author} {\bibinfo {author} {\bibfnamefont {S.}~\bibnamefont {Gandolfi}}, \bibinfo {author} {\bibfnamefont {A.}~\bibnamefont {Lovato}}, \bibinfo {author} {\bibfnamefont {J.}~\bibnamefont {Carlson}}, \ and\ \bibinfo {author} {\bibfnamefont {K.~E.}\ \bibnamefont {Schmidt}},\ }\href {\doibase 10.1103/PhysRevC.90.061306} {\bibfield  {journal} {\bibinfo  {journal} {Phys. Rev. C}\ }\textbf {\bibinfo {volume} {90}},\ \bibinfo {pages} {061306} (\bibinfo {year} {2014})}\BibitemShut {NoStop}%
\bibitem [{\citenamefont {Marino}\ \emph {et~al.}(2021)\citenamefont {Marino}, \citenamefont {Barbieri}, \citenamefont {Carbone}, \citenamefont {Col\`o}, \citenamefont {Lovato}, \citenamefont {Pederiva}, \citenamefont {Roca-Maza},\ and\ \citenamefont {Vigezzi}}]{Marino2021}%
  \BibitemOpen
  \bibfield  {author} {\bibinfo {author} {\bibfnamefont {F.}~\bibnamefont {Marino}}, \bibinfo {author} {\bibfnamefont {C.}~\bibnamefont {Barbieri}}, \bibinfo {author} {\bibfnamefont {A.}~\bibnamefont {Carbone}}, \bibinfo {author} {\bibfnamefont {G.}~\bibnamefont {Col\`o}}, \bibinfo {author} {\bibfnamefont {A.}~\bibnamefont {Lovato}}, \bibinfo {author} {\bibfnamefont {F.}~\bibnamefont {Pederiva}}, \bibinfo {author} {\bibfnamefont {X.}~\bibnamefont {Roca-Maza}}, \ and\ \bibinfo {author} {\bibfnamefont {E.}~\bibnamefont {Vigezzi}},\ }\href {\doibase 10.1103/PhysRevC.104.024315} {\bibfield  {journal} {\bibinfo  {journal} {Phys. Rev. C}\ }\textbf {\bibinfo {volume} {104}},\ \bibinfo {pages} {024315} (\bibinfo {year} {2021})}\BibitemShut {NoStop}%
\bibitem [{\citenamefont {Shavitt}\ and\ \citenamefont {Bartlett}(2009)}]{ShavittBartlett}%
  \BibitemOpen
  \bibfield  {author} {\bibinfo {author} {\bibfnamefont {I.}~\bibnamefont {Shavitt}}\ and\ \bibinfo {author} {\bibfnamefont {R.~J.}\ \bibnamefont {Bartlett}},\ }\href {\doibase 10.1017/CBO9780511596834} {\emph {\bibinfo {title} {Many-Body Methods in Chemistry and Physics: MBPT and Coupled-Cluster Theory}}},\ Cambridge Molecular Science\ (\bibinfo  {publisher} {Cambridge University Press},\ \bibinfo {year} {2009})\BibitemShut {NoStop}%
\bibitem [{\citenamefont {Bartlett}\ and\ \citenamefont {Musia\l{}}(2007)}]{Bartlett2007}%
  \BibitemOpen
  \bibfield  {author} {\bibinfo {author} {\bibfnamefont {R.~J.}\ \bibnamefont {Bartlett}}\ and\ \bibinfo {author} {\bibfnamefont {M.}~\bibnamefont {Musia\l{}}},\ }\href {\doibase 10.1103/RevModPhys.79.291} {\bibfield  {journal} {\bibinfo  {journal} {Rev. Mod. Phys.}\ }\textbf {\bibinfo {volume} {79}},\ \bibinfo {pages} {291} (\bibinfo {year} {2007})}\BibitemShut {NoStop}%
\bibitem [{\citenamefont {Hagen}\ \emph {et~al.}(2014{\natexlab{a}})\citenamefont {Hagen}, \citenamefont {Papenbrock}, \citenamefont {Hjorth-Jensen},\ and\ \citenamefont {Dean}}]{Hagen2014Review}%
  \BibitemOpen
  \bibfield  {author} {\bibinfo {author} {\bibfnamefont {G.}~\bibnamefont {Hagen}}, \bibinfo {author} {\bibfnamefont {T.}~\bibnamefont {Papenbrock}}, \bibinfo {author} {\bibfnamefont {M.}~\bibnamefont {Hjorth-Jensen}}, \ and\ \bibinfo {author} {\bibfnamefont {D.~J.}\ \bibnamefont {Dean}},\ }\href {\doibase 10.1088/0034-4885/77/9/096302} {\bibfield  {journal} {\bibinfo  {journal} {Reports on Progress in Physics}\ }\textbf {\bibinfo {volume} {77}},\ \bibinfo {pages} {096302} (\bibinfo {year} {2014}{\natexlab{a}})}\BibitemShut {NoStop}%
\bibitem [{\citenamefont {Hagen}\ \emph {et~al.}(2014{\natexlab{b}})\citenamefont {Hagen}, \citenamefont {Papenbrock}, \citenamefont {Ekstr\"om}, \citenamefont {Wendt}, \citenamefont {Baardsen}, \citenamefont {Gandolfi}, \citenamefont {Hjorth-Jensen},\ and\ \citenamefont {Horowitz}}]{Hagen2014}%
  \BibitemOpen
  \bibfield  {author} {\bibinfo {author} {\bibfnamefont {G.}~\bibnamefont {Hagen}}, \bibinfo {author} {\bibfnamefont {T.}~\bibnamefont {Papenbrock}}, \bibinfo {author} {\bibfnamefont {A.}~\bibnamefont {Ekstr\"om}}, \bibinfo {author} {\bibfnamefont {K.~A.}\ \bibnamefont {Wendt}}, \bibinfo {author} {\bibfnamefont {G.}~\bibnamefont {Baardsen}}, \bibinfo {author} {\bibfnamefont {S.}~\bibnamefont {Gandolfi}}, \bibinfo {author} {\bibfnamefont {M.}~\bibnamefont {Hjorth-Jensen}}, \ and\ \bibinfo {author} {\bibfnamefont {C.~J.}\ \bibnamefont {Horowitz}},\ }\href {\doibase 10.1103/PhysRevC.89.014319} {\bibfield  {journal} {\bibinfo  {journal} {Phys. Rev. C}\ }\textbf {\bibinfo {volume} {89}},\ \bibinfo {pages} {014319} (\bibinfo {year} {2014}{\natexlab{b}})}\BibitemShut {NoStop}%
\bibitem [{\citenamefont {Baardsen}\ \emph {et~al.}(2013)\citenamefont {Baardsen}, \citenamefont {Ekstr\"om}, \citenamefont {Hagen},\ and\ \citenamefont {Hjorth-Jensen}}]{Baardsen2013}%
  \BibitemOpen
  \bibfield  {author} {\bibinfo {author} {\bibfnamefont {G.}~\bibnamefont {Baardsen}}, \bibinfo {author} {\bibfnamefont {A.}~\bibnamefont {Ekstr\"om}}, \bibinfo {author} {\bibfnamefont {G.}~\bibnamefont {Hagen}}, \ and\ \bibinfo {author} {\bibfnamefont {M.}~\bibnamefont {Hjorth-Jensen}},\ }\href {\doibase 10.1103/PhysRevC.88.054312} {\bibfield  {journal} {\bibinfo  {journal} {Phys. Rev. C}\ }\textbf {\bibinfo {volume} {88}},\ \bibinfo {pages} {054312} (\bibinfo {year} {2013})}\BibitemShut {NoStop}%
\bibitem [{\citenamefont {Ekstr\"om}\ \emph {et~al.}(2013)\citenamefont {Ekstr\"om}, \citenamefont {Baardsen}, \citenamefont {Forss\'en}, \citenamefont {Hagen}, \citenamefont {Hjorth-Jensen}, \citenamefont {Jansen}, \citenamefont {Machleidt}, \citenamefont {Nazarewicz}, \citenamefont {Papenbrock}, \citenamefont {Sarich},\ and\ \citenamefont {Wild}}]{NNLOopt}%
  \BibitemOpen
  \bibfield  {author} {\bibinfo {author} {\bibfnamefont {A.}~\bibnamefont {Ekstr\"om}}, \bibinfo {author} {\bibfnamefont {G.}~\bibnamefont {Baardsen}}, \bibinfo {author} {\bibfnamefont {C.}~\bibnamefont {Forss\'en}}, \bibinfo {author} {\bibfnamefont {G.}~\bibnamefont {Hagen}}, \bibinfo {author} {\bibfnamefont {M.}~\bibnamefont {Hjorth-Jensen}}, \bibinfo {author} {\bibfnamefont {G.~R.}\ \bibnamefont {Jansen}}, \bibinfo {author} {\bibfnamefont {R.}~\bibnamefont {Machleidt}}, \bibinfo {author} {\bibfnamefont {W.}~\bibnamefont {Nazarewicz}}, \bibinfo {author} {\bibfnamefont {T.}~\bibnamefont {Papenbrock}}, \bibinfo {author} {\bibfnamefont {J.}~\bibnamefont {Sarich}}, \ and\ \bibinfo {author} {\bibfnamefont {S.~M.}\ \bibnamefont {Wild}},\ }\href {\doibase 10.1103/PhysRevLett.110.192502} {\bibfield  {journal} {\bibinfo  {journal} {Phys. Rev. Lett.}\ }\textbf {\bibinfo {volume} {110}},\ \bibinfo {pages} {192502} (\bibinfo {year} {2013})}\BibitemShut {NoStop}%
\bibitem [{\citenamefont {Ekstr\"om}\ \emph {et~al.}(2018)\citenamefont {Ekstr\"om}, \citenamefont {Hagen}, \citenamefont {Morris}, \citenamefont {Papenbrock},\ and\ \citenamefont {Schwartz}}]{Ekstrom2018Delta}%
  \BibitemOpen
  \bibfield  {author} {\bibinfo {author} {\bibfnamefont {A.}~\bibnamefont {Ekstr\"om}}, \bibinfo {author} {\bibfnamefont {G.}~\bibnamefont {Hagen}}, \bibinfo {author} {\bibfnamefont {T.~D.}\ \bibnamefont {Morris}}, \bibinfo {author} {\bibfnamefont {T.}~\bibnamefont {Papenbrock}}, \ and\ \bibinfo {author} {\bibfnamefont {P.~D.}\ \bibnamefont {Schwartz}},\ }\href {\doibase 10.1103/PhysRevC.97.024332} {\bibfield  {journal} {\bibinfo  {journal} {Phys. Rev. C}\ }\textbf {\bibinfo {volume} {97}},\ \bibinfo {pages} {024332} (\bibinfo {year} {2018})}\BibitemShut {NoStop}%
\bibitem [{\citenamefont {Somà}(2020)}]{Soma2020}%
  \BibitemOpen
  \bibfield  {author} {\bibinfo {author} {\bibfnamefont {V.}~\bibnamefont {Somà}},\ }\href {\doibase 10.3389/fphy.2020.00340} {\bibfield  {journal} {\bibinfo  {journal} {Frontiers in Physics}\ }\textbf {\bibinfo {volume} {8}},\ \bibinfo {pages} {340} (\bibinfo {year} {2020})}\BibitemShut {NoStop}%
\bibitem [{\citenamefont {Dickhoff}\ and\ \citenamefont {Barbieri}(2004)}]{Barbieri2004}%
  \BibitemOpen
  \bibfield  {author} {\bibinfo {author} {\bibfnamefont {W.}~\bibnamefont {Dickhoff}}\ and\ \bibinfo {author} {\bibfnamefont {C.}~\bibnamefont {Barbieri}},\ }\href {\doibase https://doi.org/10.1016/j.ppnp.2004.02.038} {\bibfield  {journal} {\bibinfo  {journal} {Progress in Particle and Nuclear Physics}\ }\textbf {\bibinfo {volume} {52}},\ \bibinfo {pages} {377} (\bibinfo {year} {2004})}\BibitemShut {NoStop}%
\bibitem [{\citenamefont {Schirmer}(2018)}]{Schirmer2018}%
  \BibitemOpen
  \bibfield  {author} {\bibinfo {author} {\bibfnamefont {J.}~\bibnamefont {Schirmer}},\ }\href {\doibase 10.1007/978-3-319-93602-4} {\emph {\bibinfo {title} {Many-Body Methods for Atoms, Molecules and Clusters}}},\ Lecture Notes in Physics\ (\bibinfo  {publisher} {Springer International Publishing},\ \bibinfo {year} {2018})\BibitemShut {NoStop}%
\bibitem [{\citenamefont {Raimondi}\ and\ \citenamefont {Barbieri}(2018)}]{Raimondi2017}%
  \BibitemOpen
  \bibfield  {author} {\bibinfo {author} {\bibfnamefont {F.}~\bibnamefont {Raimondi}}\ and\ \bibinfo {author} {\bibfnamefont {C.}~\bibnamefont {Barbieri}},\ }\href {\doibase 10.1103/PhysRevC.97.054308} {\bibfield  {journal} {\bibinfo  {journal} {Phys. Rev. C}\ }\textbf {\bibinfo {volume} {97}},\ \bibinfo {pages} {054308} (\bibinfo {year} {2018})},\ \Eprint {http://arxiv.org/abs/1709.04330} {arXiv:1709.04330 [nucl-th]} \BibitemShut {NoStop}%
\bibitem [{\citenamefont {Barbieri}\ \emph {et~al.}(2022)\citenamefont {Barbieri}, \citenamefont {Duguet},\ and\ \citenamefont {Som\`a}}]{Barbieri2022Gorkov}%
  \BibitemOpen
  \bibfield  {author} {\bibinfo {author} {\bibfnamefont {C.}~\bibnamefont {Barbieri}}, \bibinfo {author} {\bibfnamefont {T.}~\bibnamefont {Duguet}}, \ and\ \bibinfo {author} {\bibfnamefont {V.}~\bibnamefont {Som\`a}},\ }\href {\doibase 10.1103/PhysRevC.105.044330} {\bibfield  {journal} {\bibinfo  {journal} {Phys. Rev. C}\ }\textbf {\bibinfo {volume} {105}},\ \bibinfo {pages} {044330} (\bibinfo {year} {2022})}\BibitemShut {NoStop}%
\bibitem [{\citenamefont {Som\`a}\ \emph {et~al.}(2020)\citenamefont {Som\`a}, \citenamefont {Navr\'atil}, \citenamefont {Raimondi}, \citenamefont {Barbieri},\ and\ \citenamefont {Duguet}}]{Soma2020Chiral}%
  \BibitemOpen
  \bibfield  {author} {\bibinfo {author} {\bibfnamefont {V.}~\bibnamefont {Som\`a}}, \bibinfo {author} {\bibfnamefont {P.}~\bibnamefont {Navr\'atil}}, \bibinfo {author} {\bibfnamefont {F.}~\bibnamefont {Raimondi}}, \bibinfo {author} {\bibfnamefont {C.}~\bibnamefont {Barbieri}}, \ and\ \bibinfo {author} {\bibfnamefont {T.}~\bibnamefont {Duguet}},\ }\href {https://link.aps.org/doi/10.1103/PhysRevC.101.014318} {\bibfield  {journal} {\bibinfo  {journal} {Phys. Rev. C}\ }\textbf {\bibinfo {volume} {101}},\ \bibinfo {pages} {014318} (\bibinfo {year} {2020})}\BibitemShut {NoStop}%
\bibitem [{\citenamefont {Arthuis}\ \emph {et~al.}(2020)\citenamefont {Arthuis}, \citenamefont {Barbieri}, \citenamefont {Vorabbi},\ and\ \citenamefont {Finelli}}]{Arthuis_2020}%
  \BibitemOpen
  \bibfield  {author} {\bibinfo {author} {\bibfnamefont {P.}~\bibnamefont {Arthuis}}, \bibinfo {author} {\bibfnamefont {C.}~\bibnamefont {Barbieri}}, \bibinfo {author} {\bibfnamefont {M.}~\bibnamefont {Vorabbi}}, \ and\ \bibinfo {author} {\bibfnamefont {P.}~\bibnamefont {Finelli}},\ }\href {http://dx.doi.org/10.1103/PhysRevLett.125.182501} {\bibfield  {journal} {\bibinfo  {journal} {Physical Review Letters}\ }\textbf {\bibinfo {volume} {125}} (\bibinfo {year} {2020})}\BibitemShut {NoStop}%
\bibitem [{\citenamefont {Som\`a}\ \emph {et~al.}(2014)\citenamefont {Som\`a}, \citenamefont {Cipollone}, \citenamefont {Barbieri}, \citenamefont {Navr\'atil},\ and\ \citenamefont {Duguet}}]{Soma2014Chains}%
  \BibitemOpen
  \bibfield  {author} {\bibinfo {author} {\bibfnamefont {V.}~\bibnamefont {Som\`a}}, \bibinfo {author} {\bibfnamefont {A.}~\bibnamefont {Cipollone}}, \bibinfo {author} {\bibfnamefont {C.}~\bibnamefont {Barbieri}}, \bibinfo {author} {\bibfnamefont {P.}~\bibnamefont {Navr\'atil}}, \ and\ \bibinfo {author} {\bibfnamefont {T.}~\bibnamefont {Duguet}},\ }\href {\doibase 10.1103/PhysRevC.89.061301} {\bibfield  {journal} {\bibinfo  {journal} {Phys. Rev. C}\ }\textbf {\bibinfo {volume} {89}},\ \bibinfo {pages} {061301} (\bibinfo {year} {2014})}\BibitemShut {NoStop}%
\bibitem [{\citenamefont {Raimondi}\ and\ \citenamefont {Barbieri}(2019)}]{Raimondi2019}%
  \BibitemOpen
  \bibfield  {author} {\bibinfo {author} {\bibfnamefont {F.}~\bibnamefont {Raimondi}}\ and\ \bibinfo {author} {\bibfnamefont {C.}~\bibnamefont {Barbieri}},\ }\href {\doibase 10.1103/PhysRevC.99.054327} {\bibfield  {journal} {\bibinfo  {journal} {Phys. Rev. C}\ }\textbf {\bibinfo {volume} {99}},\ \bibinfo {pages} {054327} (\bibinfo {year} {2019})}\BibitemShut {NoStop}%
\bibitem [{\citenamefont {Barbieri}\ and\ \citenamefont {Carbone}(2017)}]{Barbieri2017}%
  \BibitemOpen
  \bibfield  {author} {\bibinfo {author} {\bibfnamefont {C.}~\bibnamefont {Barbieri}}\ and\ \bibinfo {author} {\bibfnamefont {A.}~\bibnamefont {Carbone}},\ }\enquote {\bibinfo {title} {Self-consistent green's function approaches},}\ in\ \href {\doibase 10.1007/978-3-319-53336-0_11} {\emph {\bibinfo {booktitle} {An Advanced Course in Computational Nuclear Physics: Bridging the Scales from Quarks to Neutron Stars}}},\ \bibinfo {editor} {edited by\ \bibinfo {editor} {\bibfnamefont {M.}~\bibnamefont {Hjorth-Jensen}}, \bibinfo {editor} {\bibfnamefont {M.~P.}\ \bibnamefont {Lombardo}}, \ and\ \bibinfo {editor} {\bibfnamefont {U.}~\bibnamefont {van Kolck}}}\ (\bibinfo  {publisher} {Springer International Publishing},\ \bibinfo {address} {Cham},\ \bibinfo {year} {2017})\ pp.\ \bibinfo {pages} {571--644}\BibitemShut {NoStop}%
\bibitem [{\citenamefont {Mcilroy}(2020)}]{McilroyChristopher2020Sgfs}%
  \BibitemOpen
  \bibfield  {author} {\bibinfo {author} {\bibfnamefont {C.}~\bibnamefont {Mcilroy}},\ }\emph {\bibinfo {title} {Self-consistent green’s function studies of modern hamiltonians in finite and infinite systems.}},\ \href@noop {} {Ph.D. thesis},\ \bibinfo  {school} {University of Surrey} (\bibinfo {year} {2020})\BibitemShut {NoStop}%
\bibitem [{\citenamefont {Marino}\ \emph {et~al.}(2024{\natexlab{a}})\citenamefont {Marino}, \citenamefont {Barbieri},\ and\ \citenamefont {Col\`{o}}}]{Marino2024}%
  \BibitemOpen
  \bibfield  {author} {\bibinfo {author} {\bibfnamefont {F.}~\bibnamefont {Marino}}, \bibinfo {author} {\bibfnamefont {C.}~\bibnamefont {Barbieri}}, \ and\ \bibinfo {author} {\bibfnamefont {G.}~\bibnamefont {Col\`{o}}},\ }\href@noop {} {\enquote {\bibinfo {title} {in preparation},}\ } (\bibinfo {year} {2024}{\natexlab{a}})\BibitemShut {NoStop}%
\bibitem [{\citenamefont {Marino}(2023)}]{MarinoPhdThesis}%
  \BibitemOpen
  \bibfield  {author} {\bibinfo {author} {\bibfnamefont {F.}~\bibnamefont {Marino}},\ }\emph {\bibinfo {title} {Microscopic theory of infinite nuclear matter and non-empirical energy functionals}},\ \href {https://hdl.handle.net/2434/1023132} {Ph.D. thesis},\ \bibinfo  {school} {Universit\`{a} degli Studi di Milano} (\bibinfo {year} {2023})\BibitemShut {NoStop}%
\bibitem [{\citenamefont {Marino}\ \emph {et~al.}(2024{\natexlab{b}})\citenamefont {Marino}, \citenamefont {Barbieri}, \citenamefont {Colò}, \citenamefont {Jiang},\ and\ \citenamefont {Novario}}]{Marino2024Qnp}%
  \BibitemOpen
  \bibfield  {author} {\bibinfo {author} {\bibfnamefont {F.}~\bibnamefont {Marino}}, \bibinfo {author} {\bibfnamefont {C.}~\bibnamefont {Barbieri}}, \bibinfo {author} {\bibfnamefont {G.}~\bibnamefont {Colò}}, \bibinfo {author} {\bibfnamefont {W.}~\bibnamefont {Jiang}}, \ and\ \bibinfo {author} {\bibfnamefont {S.~J.}\ \bibnamefont {Novario}},\ }\href {https://arxiv.org/abs/2409.07432} {\  (\bibinfo {year} {2024}{\natexlab{b}})},\ \bibinfo {note} {arxiv:2409.07432},\ \Eprint {http://arxiv.org/abs/2409.07432} {arXiv:2409.07432 [nucl-th]} \BibitemShut {NoStop}%
\bibitem [{\citenamefont {Baldo}\ \emph {et~al.}(2012)\citenamefont {Baldo}, \citenamefont {Polls}, \citenamefont {Rios}, \citenamefont {Schulze},\ and\ \citenamefont {Vida\~na}}]{Baldo2012}%
  \BibitemOpen
  \bibfield  {author} {\bibinfo {author} {\bibfnamefont {M.}~\bibnamefont {Baldo}}, \bibinfo {author} {\bibfnamefont {A.}~\bibnamefont {Polls}}, \bibinfo {author} {\bibfnamefont {A.}~\bibnamefont {Rios}}, \bibinfo {author} {\bibfnamefont {H.-J.}\ \bibnamefont {Schulze}}, \ and\ \bibinfo {author} {\bibfnamefont {I.}~\bibnamefont {Vida\~na}},\ }\href {\doibase 10.1103/PhysRevC.86.064001} {\bibfield  {journal} {\bibinfo  {journal} {Phys. Rev. C}\ }\textbf {\bibinfo {volume} {86}},\ \bibinfo {pages} {064001} (\bibinfo {year} {2012})}\BibitemShut {NoStop}%
\bibitem [{\citenamefont {Hergert}\ \emph {et~al.}(2017)\citenamefont {Hergert}, \citenamefont {Bogner}, \citenamefont {Lietz}, \citenamefont {Morris}, \citenamefont {Novario}, \citenamefont {Parzuchowski},\ and\ \citenamefont {Yuan}}]{Hergert2016}%
  \BibitemOpen
  \bibfield  {author} {\bibinfo {author} {\bibfnamefont {H.}~\bibnamefont {Hergert}}, \bibinfo {author} {\bibfnamefont {S.~K.}\ \bibnamefont {Bogner}}, \bibinfo {author} {\bibfnamefont {J.~G.}\ \bibnamefont {Lietz}}, \bibinfo {author} {\bibfnamefont {T.~D.}\ \bibnamefont {Morris}}, \bibinfo {author} {\bibfnamefont {S.}~\bibnamefont {Novario}}, \bibinfo {author} {\bibfnamefont {N.~M.}\ \bibnamefont {Parzuchowski}}, \ and\ \bibinfo {author} {\bibfnamefont {F.}~\bibnamefont {Yuan}},\ }\href {\doibase 10.1007/978-3-319-53336-0_10} {\bibfield  {journal} {\bibinfo  {journal} {Lect. Notes Phys.}\ }\textbf {\bibinfo {volume} {936}},\ \bibinfo {pages} {477} (\bibinfo {year} {2017})},\ \Eprint {http://arxiv.org/abs/1612.08315} {arXiv:1612.08315 [nucl-th]} \BibitemShut {NoStop}%
\bibitem [{\citenamefont {Scalesi}\ \emph {et~al.}(2024)\citenamefont {Scalesi}, \citenamefont {Duguet}, \citenamefont {Demol}, \citenamefont {Frosini}, \citenamefont {Somà},\ and\ \citenamefont {Tichai}}]{Scalesi2024}%
  \BibitemOpen
  \bibfield  {author} {\bibinfo {author} {\bibfnamefont {A.}~\bibnamefont {Scalesi}}, \bibinfo {author} {\bibfnamefont {T.}~\bibnamefont {Duguet}}, \bibinfo {author} {\bibfnamefont {P.}~\bibnamefont {Demol}}, \bibinfo {author} {\bibfnamefont {M.}~\bibnamefont {Frosini}}, \bibinfo {author} {\bibfnamefont {V.}~\bibnamefont {Somà}}, \ and\ \bibinfo {author} {\bibfnamefont {A.}~\bibnamefont {Tichai}},\ }\href@noop {} {\enquote {\bibinfo {title} {Impact of correlations on nuclear binding energies},}\ } (\bibinfo {year} {2024}),\ \Eprint {http://arxiv.org/abs/2406.03545} {arXiv:2406.03545 [nucl-th]} \BibitemShut {NoStop}%
\bibitem [{\citenamefont {Hebeler}\ and\ \citenamefont {Schwenk}(2010)}]{Hebeler2010}%
  \BibitemOpen
  \bibfield  {author} {\bibinfo {author} {\bibfnamefont {K.}~\bibnamefont {Hebeler}}\ and\ \bibinfo {author} {\bibfnamefont {A.}~\bibnamefont {Schwenk}},\ }\href {\doibase 10.1103/PhysRevC.82.014314} {\bibfield  {journal} {\bibinfo  {journal} {Phys. Rev. C}\ }\textbf {\bibinfo {volume} {82}},\ \bibinfo {pages} {014314} (\bibinfo {year} {2010})}\BibitemShut {NoStop}%
\bibitem [{\citenamefont {Marino}\ \emph {et~al.}(2023)\citenamefont {Marino}, \citenamefont {Col\`o}, \citenamefont {Roca-Maza},\ and\ \citenamefont {Vigezzi}}]{Marino2023}%
  \BibitemOpen
  \bibfield  {author} {\bibinfo {author} {\bibfnamefont {F.}~\bibnamefont {Marino}}, \bibinfo {author} {\bibfnamefont {G.}~\bibnamefont {Col\`o}}, \bibinfo {author} {\bibfnamefont {X.}~\bibnamefont {Roca-Maza}}, \ and\ \bibinfo {author} {\bibfnamefont {E.}~\bibnamefont {Vigezzi}},\ }\href {\doibase 10.1103/PhysRevC.107.044311} {\bibfield  {journal} {\bibinfo  {journal} {Phys. Rev. C}\ }\textbf {\bibinfo {volume} {107}},\ \bibinfo {pages} {044311} (\bibinfo {year} {2023})}\BibitemShut {NoStop}%
\bibitem [{\citenamefont {Lietz}\ \emph {et~al.}(2017)\citenamefont {Lietz}, \citenamefont {Novario}, \citenamefont {Jansen}, \citenamefont {Hagen},\ and\ \citenamefont {Hjorth-Jensen}}]{LietzCompNucl}%
  \BibitemOpen
  \bibfield  {author} {\bibinfo {author} {\bibfnamefont {J.~G.}\ \bibnamefont {Lietz}}, \bibinfo {author} {\bibfnamefont {S.}~\bibnamefont {Novario}}, \bibinfo {author} {\bibfnamefont {G.~R.}\ \bibnamefont {Jansen}}, \bibinfo {author} {\bibfnamefont {G.}~\bibnamefont {Hagen}}, \ and\ \bibinfo {author} {\bibfnamefont {M.}~\bibnamefont {Hjorth-Jensen}},\ }in\ \href@noop {} {{\selectlanguage {english}\emph {\bibinfo {booktitle} {An Advanced Course in Computational Nuclear Physics}}}},\ \bibinfo {series} {Lecture Notes in Physics}, Vol.\ \bibinfo {volume} {936}\ (\bibinfo  {publisher} {Springer International Publishing},\ \bibinfo {address} {Cham},\ \bibinfo {year} {2017})\ pp.\ \bibinfo {pages} {293--399}\BibitemShut {NoStop}%
\bibitem [{\citenamefont {Epelbaum}(2006)}]{Epelbaum2006Review}%
  \BibitemOpen
  \bibfield  {author} {\bibinfo {author} {\bibfnamefont {E.}~\bibnamefont {Epelbaum}},\ }\href {\doibase https://doi.org/10.1016/j.ppnp.2005.09.002} {\bibfield  {journal} {\bibinfo  {journal} {Progress in Particle and Nuclear Physics}\ }\textbf {\bibinfo {volume} {57}},\ \bibinfo {pages} {654} (\bibinfo {year} {2006})}\BibitemShut {NoStop}%
\bibitem [{\citenamefont {Machleidt}\ and\ \citenamefont {Entem}(2011)}]{Machleidt2011}%
  \BibitemOpen
  \bibfield  {author} {\bibinfo {author} {\bibfnamefont {R.}~\bibnamefont {Machleidt}}\ and\ \bibinfo {author} {\bibfnamefont {D.}~\bibnamefont {Entem}},\ }\href {\doibase https://doi.org/10.1016/j.physrep.2011.02.001} {\bibfield  {journal} {\bibinfo  {journal} {Physics Reports}\ }\textbf {\bibinfo {volume} {503}},\ \bibinfo {pages} {1} (\bibinfo {year} {2011})}\BibitemShut {NoStop}%
\bibitem [{\citenamefont {Hebeler}(2021)}]{Hebeler3nf}%
  \BibitemOpen
  \bibfield  {author} {\bibinfo {author} {\bibfnamefont {K.}~\bibnamefont {Hebeler}},\ }\href {https://www.sciencedirect.com/science/article/pii/S0370157320303409} {\bibfield  {journal} {\bibinfo  {journal} {Physics Reports}\ }\textbf {\bibinfo {volume} {890}},\ \bibinfo {pages} {1} (\bibinfo {year} {2021})}\BibitemShut {NoStop}%
\bibitem [{\citenamefont {Dj\"arv}\ \emph {et~al.}(2021)\citenamefont {Dj\"arv}, \citenamefont {Ekstr\"om}, \citenamefont {Forss\'en},\ and\ \citenamefont {Jansen}}]{Djarv:2021xjg}%
  \BibitemOpen
  \bibfield  {author} {\bibinfo {author} {\bibfnamefont {T.}~\bibnamefont {Dj\"arv}}, \bibinfo {author} {\bibfnamefont {A.}~\bibnamefont {Ekstr\"om}}, \bibinfo {author} {\bibfnamefont {C.}~\bibnamefont {Forss\'en}}, \ and\ \bibinfo {author} {\bibfnamefont {G.~R.}\ \bibnamefont {Jansen}},\ }\href {\doibase 10.1103/PhysRevC.104.024324} {\bibfield  {journal} {\bibinfo  {journal} {Phys. Rev. C}\ }\textbf {\bibinfo {volume} {104}},\ \bibinfo {pages} {024324} (\bibinfo {year} {2021})},\ \Eprint {http://arxiv.org/abs/2104.14820} {arXiv:2104.14820 [nucl-th]} \BibitemShut {NoStop}%
\bibitem [{\citenamefont {Hagen}\ \emph {et~al.}(2007)\citenamefont {Hagen}, \citenamefont {Papenbrock}, \citenamefont {Dean}, \citenamefont {Schwenk}, \citenamefont {Nogga}, \citenamefont {W\l{}och},\ and\ \citenamefont {Piecuch}}]{Hagen2007}%
  \BibitemOpen
  \bibfield  {author} {\bibinfo {author} {\bibfnamefont {G.}~\bibnamefont {Hagen}}, \bibinfo {author} {\bibfnamefont {T.}~\bibnamefont {Papenbrock}}, \bibinfo {author} {\bibfnamefont {D.~J.}\ \bibnamefont {Dean}}, \bibinfo {author} {\bibfnamefont {A.}~\bibnamefont {Schwenk}}, \bibinfo {author} {\bibfnamefont {A.}~\bibnamefont {Nogga}}, \bibinfo {author} {\bibfnamefont {M.}~\bibnamefont {W\l{}och}}, \ and\ \bibinfo {author} {\bibfnamefont {P.}~\bibnamefont {Piecuch}},\ }\href {\doibase 10.1103/PhysRevC.76.034302} {\bibfield  {journal} {\bibinfo  {journal} {Phys. Rev. C}\ }\textbf {\bibinfo {volume} {76}},\ \bibinfo {pages} {034302} (\bibinfo {year} {2007})}\BibitemShut {NoStop}%
\bibitem [{\citenamefont {Crawford}\ and\ \citenamefont {Schaefer}(2000)}]{CrawfordSchaefer2007}%
  \BibitemOpen
  \bibfield  {author} {\bibinfo {author} {\bibfnamefont {T.~D.}\ \bibnamefont {Crawford}}\ and\ \bibinfo {author} {\bibfnamefont {H.~F.}\ \bibnamefont {Schaefer}},\ }in\ \href {\doibase 10.1002/9780470125915.ch2} {{\selectlanguage {english}\emph {\bibinfo {booktitle} {Reviews in Computational Chemistry}}}},\ \bibinfo {editor} {edited by\ \bibinfo {editor} {\bibfnamefont {K.~B.}\ \bibnamefont {Lipkowitz}}\ and\ \bibinfo {editor} {\bibfnamefont {D.~B.}\ \bibnamefont {Boyd}}}\ (\bibinfo  {publisher} {John Wiley and Sons, Inc},\ \bibinfo {address} {Hoboken, NJ, USA},\ \bibinfo {year} {2000})\ p.\ \bibinfo {pages} {104}\BibitemShut {NoStop}%
\bibitem [{\citenamefont {Barbieri}\ and\ \citenamefont {Hjorth-Jensen}(2009)}]{Barbieri2009}%
  \BibitemOpen
  \bibfield  {author} {\bibinfo {author} {\bibfnamefont {C.}~\bibnamefont {Barbieri}}\ and\ \bibinfo {author} {\bibfnamefont {M.}~\bibnamefont {Hjorth-Jensen}},\ }\href {\doibase 10.1103/PhysRevC.79.064313} {\bibfield  {journal} {\bibinfo  {journal} {Phys. Rev. C}\ }\textbf {\bibinfo {volume} {79}},\ \bibinfo {pages} {064313} (\bibinfo {year} {2009})}\BibitemShut {NoStop}%
\bibitem [{\citenamefont {Som\`a}\ \emph {et~al.}(2013)\citenamefont {Som\`a}, \citenamefont {Barbieri},\ and\ \citenamefont {Duguet}}]{Soma2013}%
  \BibitemOpen
  \bibfield  {author} {\bibinfo {author} {\bibfnamefont {V.}~\bibnamefont {Som\`a}}, \bibinfo {author} {\bibfnamefont {C.}~\bibnamefont {Barbieri}}, \ and\ \bibinfo {author} {\bibfnamefont {T.}~\bibnamefont {Duguet}},\ }\href {\doibase 10.1103/PhysRevC.87.011303} {\bibfield  {journal} {\bibinfo  {journal} {Phys. Rev. C}\ }\textbf {\bibinfo {volume} {87}},\ \bibinfo {pages} {011303} (\bibinfo {year} {2013})}\BibitemShut {NoStop}%
\bibitem [{\citenamefont {Hodecker}\ \emph {et~al.}(2019)\citenamefont {Hodecker}, \citenamefont {Dempwolff}, \citenamefont {Rehn},\ and\ \citenamefont {Dreuw}}]{Hodecker2019}%
  \BibitemOpen
  \bibfield  {author} {\bibinfo {author} {\bibfnamefont {M.}~\bibnamefont {Hodecker}}, \bibinfo {author} {\bibfnamefont {A.~L.}\ \bibnamefont {Dempwolff}}, \bibinfo {author} {\bibfnamefont {D.~R.}\ \bibnamefont {Rehn}}, \ and\ \bibinfo {author} {\bibfnamefont {A.}~\bibnamefont {Dreuw}},\ }\href {\doibase 10.1063/1.5081663} {\bibfield  {journal} {\bibinfo  {journal} {The Journal of Chemical Physics}\ }\textbf {\bibinfo {volume} {150}},\ \bibinfo {pages} {174104} (\bibinfo {year} {2019})},\ \Eprint {http://arxiv.org/abs/https://pubs.aip.org/aip/jcp/article-pdf/doi/10.1063/1.5081663/13773006/174104\_1\_online.pdf} {https://pubs.aip.org/aip/jcp/article-pdf/doi/10.1063/1.5081663/13773006/174104\_1\_online.pdf} \BibitemShut {NoStop}%
\bibitem [{\citenamefont {Acharya}\ \emph {et~al.}(2023)\citenamefont {Acharya}, \citenamefont {Bacca}, \citenamefont {Bonaiti}, \citenamefont {Li~Muli},\ and\ \citenamefont {Sobczyk}}]{Acharya:2022drl}%
  \BibitemOpen
  \bibfield  {author} {\bibinfo {author} {\bibfnamefont {B.}~\bibnamefont {Acharya}}, \bibinfo {author} {\bibfnamefont {S.}~\bibnamefont {Bacca}}, \bibinfo {author} {\bibfnamefont {F.}~\bibnamefont {Bonaiti}}, \bibinfo {author} {\bibfnamefont {S.~S.}\ \bibnamefont {Li~Muli}}, \ and\ \bibinfo {author} {\bibfnamefont {J.~E.}\ \bibnamefont {Sobczyk}},\ }\href {\doibase 10.3389/fphy.2022.1066035} {\bibfield  {journal} {\bibinfo  {journal} {Front. Phys.}\ }\textbf {\bibinfo {volume} {10}},\ \bibinfo {pages} {1066035} (\bibinfo {year} {2023})},\ \Eprint {http://arxiv.org/abs/2210.04632} {arXiv:2210.04632 [nucl-th]} \BibitemShut {NoStop}%
\bibitem [{\citenamefont {Tews}\ \emph {et~al.}(2024)\citenamefont {Tews}, \citenamefont {Somasundaram}, \citenamefont {Lonardoni}, \citenamefont {G\"ottling}, \citenamefont {Seutin}, \citenamefont {Carlson}, \citenamefont {Gandolfi}, \citenamefont {Hebeler},\ and\ \citenamefont {Schwenk}}]{Tews:2024owl}%
  \BibitemOpen
  \bibfield  {author} {\bibinfo {author} {\bibfnamefont {I.}~\bibnamefont {Tews}}, \bibinfo {author} {\bibfnamefont {R.}~\bibnamefont {Somasundaram}}, \bibinfo {author} {\bibfnamefont {D.}~\bibnamefont {Lonardoni}}, \bibinfo {author} {\bibfnamefont {H.}~\bibnamefont {G\"ottling}}, \bibinfo {author} {\bibfnamefont {R.}~\bibnamefont {Seutin}}, \bibinfo {author} {\bibfnamefont {J.}~\bibnamefont {Carlson}}, \bibinfo {author} {\bibfnamefont {S.}~\bibnamefont {Gandolfi}}, \bibinfo {author} {\bibfnamefont {K.}~\bibnamefont {Hebeler}}, \ and\ \bibinfo {author} {\bibfnamefont {A.}~\bibnamefont {Schwenk}},\ }\href@noop {} {\  (\bibinfo {year} {2024})},\ \Eprint {http://arxiv.org/abs/2407.08979} {arXiv:2407.08979 [nucl-th]} \BibitemShut {NoStop}%
\bibitem [{\citenamefont {Wiringa}\ \emph {et~al.}(1995)\citenamefont {Wiringa}, \citenamefont {Stoks},\ and\ \citenamefont {Schiavilla}}]{Wiringa1995}%
  \BibitemOpen
  \bibfield  {author} {\bibinfo {author} {\bibfnamefont {R.~B.}\ \bibnamefont {Wiringa}}, \bibinfo {author} {\bibfnamefont {V.~G.~J.}\ \bibnamefont {Stoks}}, \ and\ \bibinfo {author} {\bibfnamefont {R.}~\bibnamefont {Schiavilla}},\ }\href {\doibase 10.1103/PhysRevC.51.38} {\bibfield  {journal} {\bibinfo  {journal} {Phys. Rev. C}\ }\textbf {\bibinfo {volume} {51}},\ \bibinfo {pages} {38} (\bibinfo {year} {1995})}\BibitemShut {NoStop}%
\bibitem [{\citenamefont {Wiringa}\ and\ \citenamefont {Pieper}(2002)}]{Wiringa2002}%
  \BibitemOpen
  \bibfield  {author} {\bibinfo {author} {\bibfnamefont {R.~B.}\ \bibnamefont {Wiringa}}\ and\ \bibinfo {author} {\bibfnamefont {S.~C.}\ \bibnamefont {Pieper}},\ }\href {\doibase 10.1103/PhysRevLett.89.182501} {\bibfield  {journal} {\bibinfo  {journal} {Phys. Rev. Lett.}\ }\textbf {\bibinfo {volume} {89}},\ \bibinfo {pages} {182501} (\bibinfo {year} {2002})}\BibitemShut {NoStop}%
\bibitem [{\citenamefont {Rios}\ \emph {et~al.}(2017)\citenamefont {Rios}, \citenamefont {Carbone},\ and\ \citenamefont {Polls}}]{Rios2017}%
  \BibitemOpen
  \bibfield  {author} {\bibinfo {author} {\bibfnamefont {A.}~\bibnamefont {Rios}}, \bibinfo {author} {\bibfnamefont {A.}~\bibnamefont {Carbone}}, \ and\ \bibinfo {author} {\bibfnamefont {A.}~\bibnamefont {Polls}},\ }\href {\doibase 10.1103/PhysRevC.96.014003} {\bibfield  {journal} {\bibinfo  {journal} {Phys. Rev. C}\ }\textbf {\bibinfo {volume} {96}},\ \bibinfo {pages} {014003} (\bibinfo {year} {2017})}\BibitemShut {NoStop}%
\bibitem [{\citenamefont {Mattuck}(1967)}]{Mattuck}%
  \BibitemOpen
  \bibfield  {author} {\bibinfo {author} {\bibfnamefont {R.~D.}\ \bibnamefont {Mattuck}},\ }\href@noop {} {\emph {\bibinfo {title} {A guide to {Feynman} diagrams in the many-body problem}}},\ European physics series\ (\bibinfo  {publisher} {pub-MCGRAW-HILL},\ \bibinfo {address} {pub-MCGRAW-HILL:adr},\ \bibinfo {year} {1967})\ pp.\ \bibinfo {pages} {xii + 294}\BibitemShut {NoStop}%
\bibitem [{\citenamefont {Gezerlis}\ \emph {et~al.}(2013)\citenamefont {Gezerlis}, \citenamefont {Tews}, \citenamefont {Epelbaum}, \citenamefont {Gandolfi}, \citenamefont {Hebeler}, \citenamefont {Nogga},\ and\ \citenamefont {Schwenk}}]{Gezerlis2013}%
  \BibitemOpen
  \bibfield  {author} {\bibinfo {author} {\bibfnamefont {A.}~\bibnamefont {Gezerlis}}, \bibinfo {author} {\bibfnamefont {I.}~\bibnamefont {Tews}}, \bibinfo {author} {\bibfnamefont {E.}~\bibnamefont {Epelbaum}}, \bibinfo {author} {\bibfnamefont {S.}~\bibnamefont {Gandolfi}}, \bibinfo {author} {\bibfnamefont {K.}~\bibnamefont {Hebeler}}, \bibinfo {author} {\bibfnamefont {A.}~\bibnamefont {Nogga}}, \ and\ \bibinfo {author} {\bibfnamefont {A.}~\bibnamefont {Schwenk}},\ }\href {\doibase 10.1103/PhysRevLett.111.032501} {\bibfield  {journal} {\bibinfo  {journal} {Phys. Rev. Lett.}\ }\textbf {\bibinfo {volume} {111}},\ \bibinfo {pages} {032501} (\bibinfo {year} {2013})}\BibitemShut {NoStop}%
\bibitem [{\citenamefont {Entem}\ \emph {et~al.}(2017)\citenamefont {Entem}, \citenamefont {Machleidt},\ and\ \citenamefont {Nosyk}}]{Entem2017}%
  \BibitemOpen
  \bibfield  {author} {\bibinfo {author} {\bibfnamefont {D.~R.}\ \bibnamefont {Entem}}, \bibinfo {author} {\bibfnamefont {R.}~\bibnamefont {Machleidt}}, \ and\ \bibinfo {author} {\bibfnamefont {Y.}~\bibnamefont {Nosyk}},\ }\href {\doibase 10.1103/PhysRevC.96.024004} {\bibfield  {journal} {\bibinfo  {journal} {Phys. Rev. C}\ }\textbf {\bibinfo {volume} {96}},\ \bibinfo {pages} {024004} (\bibinfo {year} {2017})}\BibitemShut {NoStop}%
\bibitem [{\citenamefont {Weikert}\ \emph {et~al.}(1996)\citenamefont {Weikert}, \citenamefont {Meyer}, \citenamefont {Cederbaum},\ and\ \citenamefont {Tarantelli}}]{Weikert}%
  \BibitemOpen
  \bibfield  {author} {\bibinfo {author} {\bibfnamefont {H.}~\bibnamefont {Weikert}}, \bibinfo {author} {\bibfnamefont {H.}~\bibnamefont {Meyer}}, \bibinfo {author} {\bibfnamefont {L.~S.}\ \bibnamefont {Cederbaum}}, \ and\ \bibinfo {author} {\bibfnamefont {F.}~\bibnamefont {Tarantelli}},\ }\href {\doibase 10.1063/1.471429} {\bibfield  {journal} {\bibinfo  {journal} {The Journal of Chemical Physics}\ }\textbf {\bibinfo {volume} {104}},\ \bibinfo {pages} {7122} (\bibinfo {year} {1996})},\ \Eprint {http://arxiv.org/abs/https://pubs.aip.org/aip/jcp/article-pdf/104/18/7122/19077044/7122\_1\_online.pdf} {https://pubs.aip.org/aip/jcp/article-pdf/104/18/7122/19077044/7122\_1\_online.pdf} \BibitemShut {NoStop}%
\bibitem [{\citenamefont {Banerjee}\ and\ \citenamefont {Sokolov}(2023)}]{Banerjee2023}%
  \BibitemOpen
  \bibfield  {author} {\bibinfo {author} {\bibfnamefont {S.}~\bibnamefont {Banerjee}}\ and\ \bibinfo {author} {\bibfnamefont {A.~Y.}\ \bibnamefont {Sokolov}},\ }\href {\doibase 10.1021/acs.jctc.3c00251} {\bibfield  {journal} {\bibinfo  {journal} {Journal of Chemical Theory and Computation}\ }\textbf {\bibinfo {volume} {19}},\ \bibinfo {pages} {3037} (\bibinfo {year} {2023})},\ \bibinfo {note} {pMID: 37191264},\ \Eprint {http://arxiv.org/abs/https://doi.org/10.1021/acs.jctc.3c00251} {https://doi.org/10.1021/acs.jctc.3c00251} \BibitemShut {NoStop}%
\bibitem [{\citenamefont {Shepherd}\ \emph {et~al.}(2014)\citenamefont {Shepherd}, \citenamefont {Henderson},\ and\ \citenamefont {Scuseria}}]{Shepherd2014}%
  \BibitemOpen
  \bibfield  {author} {\bibinfo {author} {\bibfnamefont {J.~J.}\ \bibnamefont {Shepherd}}, \bibinfo {author} {\bibfnamefont {T.~M.}\ \bibnamefont {Henderson}}, \ and\ \bibinfo {author} {\bibfnamefont {G.~E.}\ \bibnamefont {Scuseria}},\ }\href {\doibase 10.1063/1.4867783} {\bibfield  {journal} {\bibinfo  {journal} {The Journal of Chemical Physics}\ }\textbf {\bibinfo {volume} {140}},\ \bibinfo {pages} {124102} (\bibinfo {year} {2014})},\ \Eprint {http://arxiv.org/abs/https://pubs.aip.org/aip/jcp/article-pdf/doi/10.1063/1.4867783/13234690/124102\_1\_online.pdf} {https://pubs.aip.org/aip/jcp/article-pdf/doi/10.1063/1.4867783/13234690/124102\_1\_online.pdf} \BibitemShut {NoStop}%
\bibitem [{\citenamefont {Colò}(2020)}]{colo2020}%
  \BibitemOpen
  \bibfield  {author} {\bibinfo {author} {\bibfnamefont {G.}~\bibnamefont {Colò}},\ }\href {\doibase 10.1080/23746149.2020.1740061} {\bibfield  {journal} {\bibinfo  {journal} {Advances in Physics: X}\ }\textbf {\bibinfo {volume} {5}},\ \bibinfo {pages} {1740061} (\bibinfo {year} {2020})}\BibitemShut {NoStop}%
\bibitem [{\citenamefont {Schunck}(2019)}]{Schunck2019}%
  \BibitemOpen
  \bibinfo {editor} {\bibfnamefont {N.}~\bibnamefont {Schunck}},\ ed.,\ \href {\doibase 10.1088/2053-2563/aae0ed} {\emph {\bibinfo {title} {Energy Density Functional Methods for Atomic Nuclei}}},\ 2053-2563\ (\bibinfo  {publisher} {IOP Publishing},\ \bibinfo {year} {2019})\BibitemShut {NoStop}%
\bibitem [{\citenamefont {Drischler}\ \emph {et~al.}(2024)\citenamefont {Drischler}, \citenamefont {Giuliani}, \citenamefont {Bezoui}, \citenamefont {Piekarewicz},\ and\ \citenamefont {Viens}}]{Drischler2024Bayesian}%
  \BibitemOpen
  \bibfield  {author} {\bibinfo {author} {\bibfnamefont {C.}~\bibnamefont {Drischler}}, \bibinfo {author} {\bibfnamefont {P.~G.}\ \bibnamefont {Giuliani}}, \bibinfo {author} {\bibfnamefont {S.}~\bibnamefont {Bezoui}}, \bibinfo {author} {\bibfnamefont {J.}~\bibnamefont {Piekarewicz}}, \ and\ \bibinfo {author} {\bibfnamefont {F.}~\bibnamefont {Viens}},\ }\href@noop {} {\enquote {\bibinfo {title} {A bayesian mixture model approach to quantifying the empirical nuclear saturation point},}\ } (\bibinfo {year} {2024}),\ \Eprint {http://arxiv.org/abs/2405.02748} {arXiv:2405.02748 [nucl-th]} \BibitemShut {NoStop}%
\bibitem [{\citenamefont {Drischler}(2023)}]{saturationGitHub}%
  \BibitemOpen
  \bibfield  {author} {\bibinfo {author} {\bibfnamefont {C.}~\bibnamefont {Drischler}},\ }\href {\mbox{https://github.com/cdrischler/nuclear_saturation}} {\emph {\bibinfo {title} {{{Supplemental source code on GitHub}}}}} (\bibinfo {year} {2023}),\ \bibinfo {note} {\url{https://github.com/cdrischler/nuclear_saturation}}\BibitemShut {NoStop}%
\bibitem [{\citenamefont {Ekstr\"om}\ \emph {et~al.}(2024)\citenamefont {Ekstr\"om}, \citenamefont {Jansen}, \citenamefont {Wendt}, \citenamefont {Hagen}, \citenamefont {Papenbrock}, \citenamefont {Carlsson}, \citenamefont {Forss\'en}, \citenamefont {Hjorth-Jensen}, \citenamefont {Navr\'atil},\ and\ \citenamefont {Nazarewicz}}]{NNLOsatErratum}%
  \BibitemOpen
  \bibfield  {author} {\bibinfo {author} {\bibfnamefont {A.}~\bibnamefont {Ekstr\"om}}, \bibinfo {author} {\bibfnamefont {G.~R.}\ \bibnamefont {Jansen}}, \bibinfo {author} {\bibfnamefont {K.~A.}\ \bibnamefont {Wendt}}, \bibinfo {author} {\bibfnamefont {G.}~\bibnamefont {Hagen}}, \bibinfo {author} {\bibfnamefont {T.}~\bibnamefont {Papenbrock}}, \bibinfo {author} {\bibfnamefont {B.~D.}\ \bibnamefont {Carlsson}}, \bibinfo {author} {\bibfnamefont {C.}~\bibnamefont {Forss\'en}}, \bibinfo {author} {\bibfnamefont {M.}~\bibnamefont {Hjorth-Jensen}}, \bibinfo {author} {\bibfnamefont {P.}~\bibnamefont {Navr\'atil}}, \ and\ \bibinfo {author} {\bibfnamefont {W.}~\bibnamefont {Nazarewicz}},\ }\href {\doibase 10.1103/PhysRevC.109.059901} {\bibfield  {journal} {\bibinfo  {journal} {Phys. Rev. C}\ }\textbf {\bibinfo {volume} {109}},\ \bibinfo {pages} {059901(R)} (\bibinfo {year} {2024})}\BibitemShut {NoStop}%
\bibitem [{\citenamefont {{Hu}}\ \emph {et~al.}(2024)\citenamefont {{Hu}}, \citenamefont {{Jiang}}, \citenamefont {{Miyagi}}, \citenamefont {{Sun}}, \citenamefont {{Ekstr{\"o}m}}, \citenamefont {{Forss{\'e}n}}, \citenamefont {{Hagen}}, \citenamefont {{Holt}}, \citenamefont {{Papenbrock}}, \citenamefont {{Stroberg}},\ and\ \citenamefont {{Vernon}}}]{PbAbInitioAuthorCorrection}%
  \BibitemOpen
  \bibfield  {author} {\bibinfo {author} {\bibfnamefont {B.}~\bibnamefont {{Hu}}}, \bibinfo {author} {\bibfnamefont {W.}~\bibnamefont {{Jiang}}}, \bibinfo {author} {\bibfnamefont {T.}~\bibnamefont {{Miyagi}}}, \bibinfo {author} {\bibfnamefont {Z.}~\bibnamefont {{Sun}}}, \bibinfo {author} {\bibfnamefont {A.}~\bibnamefont {{Ekstr{\"o}m}}}, \bibinfo {author} {\bibfnamefont {C.}~\bibnamefont {{Forss{\'e}n}}}, \bibinfo {author} {\bibfnamefont {G.}~\bibnamefont {{Hagen}}}, \bibinfo {author} {\bibfnamefont {J.~D.}\ \bibnamefont {{Holt}}}, \bibinfo {author} {\bibfnamefont {T.}~\bibnamefont {{Papenbrock}}}, \bibinfo {author} {\bibfnamefont {S.~R.}\ \bibnamefont {{Stroberg}}}, \ and\ \bibinfo {author} {\bibfnamefont {I.}~\bibnamefont {{Vernon}}},\ }\href {\doibase 10.1038/s41567-023-02324-9} {\bibfield  {journal} {\bibinfo  {journal} {Nature Physics}\ }\textbf {\bibinfo {volume} {20}},\ \bibinfo {pages} {169} (\bibinfo {year} {2024})}\BibitemShut {NoStop}%
\bibitem [{\citenamefont {Garcia~Ruiz}\ \emph {et~al.}(2016)\citenamefont {Garcia~Ruiz}, \citenamefont {Bissell}, \citenamefont {Blaum}, \citenamefont {Ekström}, \citenamefont {Frömmgen}, \citenamefont {Hagen}, \citenamefont {Hammen}, \citenamefont {Hebeler}, \citenamefont {Holt}, \citenamefont {Jansen}, \citenamefont {Kowalska}, \citenamefont {Kreim}, \citenamefont {Nazarewicz}, \citenamefont {Neugart}, \citenamefont {Neyens}, \citenamefont {Nörtershäuser}, \citenamefont {Papenbrock}, \citenamefont {Papuga}, \citenamefont {Schwenk},\ and\ \citenamefont {Yordanov}}]{GarciaRuiz2016}%
  \BibitemOpen
  \bibfield  {author} {\bibinfo {author} {\bibfnamefont {R.}~\bibnamefont {Garcia~Ruiz}}, \bibinfo {author} {\bibfnamefont {M.}~\bibnamefont {Bissell}}, \bibinfo {author} {\bibfnamefont {K.}~\bibnamefont {Blaum}}, \bibinfo {author} {\bibfnamefont {A.}~\bibnamefont {Ekström}}, \bibinfo {author} {\bibfnamefont {N.}~\bibnamefont {Frömmgen}}, \bibinfo {author} {\bibfnamefont {G.}~\bibnamefont {Hagen}}, \bibinfo {author} {\bibfnamefont {M.}~\bibnamefont {Hammen}}, \bibinfo {author} {\bibfnamefont {K.}~\bibnamefont {Hebeler}}, \bibinfo {author} {\bibfnamefont {J.}~\bibnamefont {Holt}}, \bibinfo {author} {\bibfnamefont {G.}~\bibnamefont {Jansen}}, \bibinfo {author} {\bibfnamefont {M.}~\bibnamefont {Kowalska}}, \bibinfo {author} {\bibfnamefont {K.}~\bibnamefont {Kreim}}, \bibinfo {author} {\bibfnamefont {W.}~\bibnamefont {Nazarewicz}}, \bibinfo {author} {\bibfnamefont {R.}~\bibnamefont {Neugart}}, \bibinfo {author} {\bibfnamefont {G.}~\bibnamefont {Neyens}}, \bibinfo {author} {\bibfnamefont {W.}~\bibnamefont
  {Nörtershäuser}}, \bibinfo {author} {\bibfnamefont {T.}~\bibnamefont {Papenbrock}}, \bibinfo {author} {\bibfnamefont {J.}~\bibnamefont {Papuga}}, \bibinfo {author} {\bibfnamefont {A.}~\bibnamefont {Schwenk}}, \ and\ \bibinfo {author} {\bibfnamefont {D.}~\bibnamefont {Yordanov}},\ }\href {\doibase 10.1038/nphys3645} {\bibfield  {journal} {\bibinfo  {journal} {Nature Physics}\ }\textbf {\bibinfo {volume} {12}},\ \bibinfo {pages} {594} (\bibinfo {year} {2016})}\BibitemShut {NoStop}%
\bibitem [{\citenamefont {Koszor\'{u}s}\ \emph {et~al.}(2021)\citenamefont {Koszor\'{u}s}, \citenamefont {Yang}, \citenamefont {Jiang}, \citenamefont {Novario}, \citenamefont {Bai}, \citenamefont {Billowes}, \citenamefont {Binnersley}, \citenamefont {Bissell}, \citenamefont {Cocolios}, \citenamefont {Cooper}, \citenamefont {de~Groote}, \citenamefont {Ekström}, \citenamefont {Flanagan}, \citenamefont {Forssén}, \citenamefont {Franchoo}, \citenamefont {Ruiz}, \citenamefont {Gustafsson}, \citenamefont {Hagen}, \citenamefont {Jansen}, \citenamefont {Kanellakopoulos}, \citenamefont {Kortelainen}, \citenamefont {Nazarewicz}, \citenamefont {Neyens}, \citenamefont {Papenbrock}, \citenamefont {Reinhard}, \citenamefont {Ricketts}, \citenamefont {Sahoo}, \citenamefont {Vernon},\ and\ \citenamefont {Wilkins}}]{Koszorus2020}%
  \BibitemOpen
  \bibfield  {author} {\bibinfo {author} {\bibfnamefont {A.}~\bibnamefont {Koszor\'{u}s}}, \bibinfo {author} {\bibfnamefont {X.~F.}\ \bibnamefont {Yang}}, \bibinfo {author} {\bibfnamefont {W.~G.}\ \bibnamefont {Jiang}}, \bibinfo {author} {\bibfnamefont {S.~J.}\ \bibnamefont {Novario}}, \bibinfo {author} {\bibfnamefont {S.~W.}\ \bibnamefont {Bai}}, \bibinfo {author} {\bibfnamefont {J.}~\bibnamefont {Billowes}}, \bibinfo {author} {\bibfnamefont {C.~L.}\ \bibnamefont {Binnersley}}, \bibinfo {author} {\bibfnamefont {M.~L.}\ \bibnamefont {Bissell}}, \bibinfo {author} {\bibfnamefont {T.~E.}\ \bibnamefont {Cocolios}}, \bibinfo {author} {\bibfnamefont {B.~S.}\ \bibnamefont {Cooper}}, \bibinfo {author} {\bibfnamefont {R.~P.}\ \bibnamefont {de~Groote}}, \bibinfo {author} {\bibfnamefont {A.}~\bibnamefont {Ekström}}, \bibinfo {author} {\bibfnamefont {K.~T.}\ \bibnamefont {Flanagan}}, \bibinfo {author} {\bibfnamefont {C.}~\bibnamefont {Forssén}}, \bibinfo {author} {\bibfnamefont {S.}~\bibnamefont {Franchoo}}, \bibinfo
  {author} {\bibfnamefont {R.~G.}\ \bibnamefont {Ruiz}}, \bibinfo {author} {\bibfnamefont {F.~P.}\ \bibnamefont {Gustafsson}}, \bibinfo {author} {\bibfnamefont {G.}~\bibnamefont {Hagen}}, \bibinfo {author} {\bibfnamefont {G.~R.}\ \bibnamefont {Jansen}}, \bibinfo {author} {\bibfnamefont {A.}~\bibnamefont {Kanellakopoulos}}, \bibinfo {author} {\bibfnamefont {M.}~\bibnamefont {Kortelainen}}, \bibinfo {author} {\bibfnamefont {W.}~\bibnamefont {Nazarewicz}}, \bibinfo {author} {\bibfnamefont {G.}~\bibnamefont {Neyens}}, \bibinfo {author} {\bibfnamefont {T.}~\bibnamefont {Papenbrock}}, \bibinfo {author} {\bibfnamefont {P.-G.}\ \bibnamefont {Reinhard}}, \bibinfo {author} {\bibfnamefont {C.~M.}\ \bibnamefont {Ricketts}}, \bibinfo {author} {\bibfnamefont {B.~K.}\ \bibnamefont {Sahoo}}, \bibinfo {author} {\bibfnamefont {A.~R.}\ \bibnamefont {Vernon}}, \ and\ \bibinfo {author} {\bibfnamefont {S.~G.}\ \bibnamefont {Wilkins}},\ }\href {\doibase 10.1038/s41567-020-01136-5} {\bibfield  {journal} {\bibinfo  {journal} {Nature
  Physics}\ }\textbf {\bibinfo {volume} {17}} (\bibinfo {year} {2021}),\ 10.1038/s41567-020-01136-5}\BibitemShut {NoStop}%
\end{thebibliography}%

\end{document}